\PassOptionsToPackage{table}{xcolor}
\documentclass[conference]{IEEEtran-NDSS}

\ifCLASSINFOpdf
\else
\fi

\let\labelindent\relax
\usepackage{enumitem}
\usepackage{hyperref}
\usepackage{graphicx}

\usepackage{listings}
\usepackage{minted}

\usepackage{array}
\usepackage{multirow}
\usepackage{ragged2e}
\usepackage{multicol}
\usepackage{amsmath} 
\usepackage{color, colortbl}

\usepackage{tikz}
\usepackage{amsmath}
\usepackage{colortbl}

\usepackage{multirow}
\usepackage{tabularx}
\usepackage[linesnumbered,ruled,vlined]{algorithm2e}
\usepackage{makecell}
\usepackage{booktabs}
\usepackage{soul}
\usepackage{changepage}
\usepackage{threeparttable}
\usepackage[table]{xcolor}

\usepackage{textcomp}
\usepackage{bbding}
\usepackage{wasysym}
\usepackage{amssymb}
\usepackage{comment}
\usepackage{makecell}
\usepackage{mfirstuc}
\usepackage{xurl}
\usepackage{pifont}
\usepackage{subcaption}

\usepackage{makecell}

\usepackage[available,functional]{ndssbadges}

\newcommand{\cc}[1]{\mbox{\smaller[0.5]\texttt{#1}}}
\newcommand{\etal}{{\em et al.}\xspace}
\newcommand{\eg}{{\em e.g.,}\xspace}
\newcommand{\ie}{{\em i.e.,}\xspace}

\newcommand{\PP}[1]{\vspace{2px}\noindent{\bf#1.}\xspace}

\newcommand*\WC[1]{%
	\begin{tikzpicture}[baseline=(C.base)]
		\node[draw,circle,inner sep=0.2pt](C) {#1};
\end{tikzpicture}}

\hypersetup{
    colorlinks=true,
    linkcolor=blue,   
}

\newcommand{\UP}[1]{\textcolor{black}{#1}} %

\begin{document}

\title{A Deep Dive into Function Inlining and its \\ Security Implications for ML-based Binary Analysis}

\author{\IEEEauthorblockN{Anonymous Authors}}

\author{
\IEEEauthorblockN{
Omar Abusabha,
Jiyong Uhm,
Tamer Abuhmed,
Hyungjoon Koo\IEEEauthorrefmark{1}~\thanks{~\IEEEauthorrefmark{1}~Corresponding author.}
}
\IEEEauthorblockA{Sungkyunkwan University, South Korea\\
404970@g.skku.edu,
jiyong423@g.skku.edu,
tamer@skku.edu,
kevin.koo@skku.edu
}
}

\IEEEoverridecommandlockouts
\makeatletter\def\@IEEEpubidpullup{6.5\baselineskip}\makeatother
\IEEEpubid{
\parbox{\columnwidth}{
Network and Distributed System Security (NDSS) Symposium 2026\\
23–27 February 2026, San Diego, CA, USA\\
ISBN 979-8-9919276-8-0\\
\url{https://dx.doi.org/10.14722/ndss.2026.241872}\\
\url{www.ndss-symposium.org}
}
\hspace{\columnsep}\makebox[\columnwidth]{}
}

\maketitle

\begin{abstract}

A function inlining optimization is 
a widely used transformation in
modern compilers, which replaces a call site
with the callee's body in need. 
While this transformation improves 
performance, it significantly impacts 
static features such as 
machine instructions 
and control flow graphs, which are
crucial to binary analysis.
Yet, despite its broad impact, 
the security impact of function inlining 
remains underexplored to date.
In this paper, we present 
the first comprehensive study of
function inlining through the lens
of machine learning-based binary analysis.
To this end, we
dissect the inlining decision pipeline
within the LLVM's cost model
and explore the combinations of the compiler
options that aggressively promote the function
inlining ratio beyond standard optimization
levels, which we term \emph{extreme inlining}.
We focus on five ML-assisted 
\UP{binary analysis tasks for security}, 
using \UP{20} unique models to systematically 
evaluate their robustness under 
extreme inlining scenarios.
Our extensive experiments reveal 
several significant findings:
i)~function inlining,
though a benign transformation in intent, 
can (in)directly affect 
ML model behaviors, being 
potentially exploited by
evading discriminative or generative 
ML models;
ii)~ML models relying on static features 
can be highly sensitive to inlining;
iii)~subtle compiler settings can 
be leveraged to deliberately craft evasive 
binary variants; and
iv)~inlining ratios vary substantially across 
applications and build configurations,
undermining assumptions of consistency
in training and evaluation of ML models.

\end{abstract}

\IEEEpeerreviewmaketitle

\section{Introduction}

Today, most applications are distributed 
in the form of executable binaries 
containing low-level machine instructions 
and embedded data.
Modern compilers produce these binaries
via a sophisticated compilation pipeline,
which involves a broad spectrum of 
optimizations.
Among them, function inlining is 
a well-established optimization technique 
that replaces a function call site 
with the body of the callee.
While this may appear to be 
a trivial transformation, determining an 
optimal inlining strategy is 
challenging -- comparable 
to the NP-complete 
Knapsack problem~\cite{inline_sub}
in complexity.

When the source code is unavailable, 
binary reverse engineering (\ie binary reversing) 
is commonly employed to comprehend 
the inner workings of a binary, unveiling
hidden or unknown functionality.
The binary reversing can be performed either 
statically~\cite{xue2019machine, binkit,
dl_fhmc, android_mal} 
(\ie analysis without executing a binary)
or dynamically~\cite{egele2014blanket, ramos2015under, yun2018qsym}.
Since the disassembled instructions in 
binaries lack high-level information 
(\eg function names, variable names, 
or types), traditional approaches extract 
a wide range of features, such as
numeric patterns, control and data flows, 
and instruction sequences.
Such features play a crucial role in
security tasks, including function boundary 
detection~\cite{BYTEWEIGHT, XDA},
binary code similarity detection 
(BCSD)~\cite{asm2vec, binshot, pei2020trex, 
safe, xu2017neural, wang2022jtrans}, vulnerability 
detection~\cite{vulseeker, shirani2018b}, 
malware analysis~\cite{android_mal}, 
malware family classification~\cite{ml_detection_family_hyper,
genealogy_malware}, 
and crash root cause 
analysis~\cite{park2023benzene}.

Recent advances integrate machine learning 
(ML) (\eg deep learning)
techniques to analyze binaries
in a high-dimensional feature space.
A vast amount of research in the literature 
have demonstrated the effectiveness
of this new direction, guiding 
a reverse engineer
by inferring function 
names~\cite{asmdepictor, debin, nero}, 
variable names~\cite{debin, chen2022augmenting} 
and types~\cite{debin, chen2022augmenting}, and 
even recovering decompiled 
code~\cite{fu2019coda, katz2019towards},
in addition to the tasks listed above.

However, function inlining as a 
common compiler optimization
can \emph{substantially distort static features} 
for reverse engineering. 
Inlining merges the callee’s instructions 
into the caller, 
which can drastically alter 
machine code
and control flow structures.
These effects are amplified with 
nested inlining 
or when additional optimizations follow. 
Consequently, static features 
may no longer be reliable
for decision-making in ML models
when aggressive inlining is 
applied (on purpose).

Although the substantial impact of
function inlining is well known, an 
\emph{in-depth study of its security 
implications on ML-based models} 
(\eg to what extent) has remained 
underexplored in the existing 
literature.
While several works have considered 
function inlining~\cite{1to1_1ton, Cross-Inlining, bingo, asm2vec},
others have occasionally 
misrepresented~\cite{InconvenientTruthsofGroundTruth}, underestimated~\cite{binkit, kim2022reverse, fsmell, discovRE}, 
heuristically addressed~\cite{bingo, asm2vec, yakdan2015no, yakdan2016helping} it, or
\UP{paid limited attention}~\cite{cross_arch_bug_search, 
                    discovRE, 
                    Inlining-vulunerbility, 
                    extracting_cross_platform,
                    xue2019machine, 
                    jin2022symlm, 
                    patrick2020probabilistic, 
                    katz2018using, XDA, BYTEWEIGHT}.
To the best of our knowledge, this is the first study that delves into
\WC{1} uncovering the security implications of function inlining in the context of
modern ML-based models;
\WC{2} investigating the entire decision-making
pipeline of inlining a function, including
the internal workings of the compiler's 
cost model;
\WC{3} exploring a broad set of 
(often overlooked) compiler options 
that affect function inlining decisions; and
\WC{4} identifying a particular combination 
of options that can increase 
the inlining ratio (\eg up to 79.64\% in
our experiments).
Our study is grounded in the 
LLVM compiler toolchain,
which offers well-modularized building blocks.
We leverage debugging information and
intermediate compiler-producing outputs 
to construct the ground truth.

LLVM’s inlining process, in essence,
follows a structured pipeline: the frontend 
(\ie Clang) and the middle end (\ie Opt) 
apply a series of analyses and transformation passes 
on intermediate representations. 
The CGSCC (Call-Graph Strongly Connected Components) pass manager internally 
coordinates  inlining and other 
call-graph-based optimizations 
across strongly connected components
(\ie sets of mutually reachable functions). 
The function inlining pass, as part of CGSCC, 
evaluates each function candidate with 
a cost model to compare 
a function’s inlining cost with a threshold. 
This pass is applied iteratively 
following each update to the call graph.
In the meantime, a multitude of 
compiler options 
(\autoref{tbl:flags}) 
can affect this decision.

In this work, we examine the potential
misuse of function inlining
as a means of circumventing ML-based
security models.
We identify a specific configuration
of compiler options that aggressively 
increases the inlining ratio, 
a strategy we term 
\emph{extreme inlining}.
Notably, we demonstrate how 
this otherwise legitimate compiler mechanism 
can be exploited by adversaries to introduce 
\emph{deliberate mutations via 
extreme inlining}, thereby enabling evasion
of both discriminative and generative
ML-based security models.
\UP{Our threat model assumes a standard build process 
repurposed for malicious ends without the compiler's 
modification or obfuscation.}
Additionally, \UP{we revisit varying inlining
scenarios, clarifying subtleties in 
function inlining practices.}

To validate our claims, we conduct extensive 
experiments by defining nine research 
questions from two different perspectives.
First, we explore how function inlining affects 
ML-assisted security applications with 
the following five tasks by training 
\UP{20} unique models (Section~\ref{ss:eval2}): 
\WC{1} binary code similarity detection,
\WC{2} function symbol name prediction,
\WC{3} malware detection,
\WC{4} malware family classification, and
\WC{5} vulnerability detection.
Second, we evaluate the function inlining optimization itself 
in terms of its ratio and further impacts with the following four questions (Section~\ref{eval1}):
\WC{6} function inlining ratio and optimization levels 
across different applications,
\WC{7} the ratio and the (frontend) compiler options,
\WC{8} the ratio and a combination of 
(middle end) compiler options 
toward extreme inlining, and
\WC{9} the variation of 
static features in a binary
according to (extreme) inlining.

Our key findings indicate that
\WC{1} function inlining (itself), 
while intended 
as a benign optimization, 
can be exploited by an adversary to 
evade both discriminative 
and generative ML models, particularly
those that rely on static features 
and are sensitive to such 
transformations;
\WC{2} subtle inlining configurations 
that affect a decision pipeline
can be deliberately manipulated to 
produce evasive binary variants
with ease; and
\WC{3} inlining ratios can vary
significantly depending on 
application-specific factors 
(\eg programming style) 
and build settings  
(\eg optimization levels, compiler options), offering ample flexibility 
for generating diverse mutations.

The following summarizes our contributions.
\begin{itemize}[leftmargin=*]
	\item 
	To the best of our knowledge, this is the first comprehensive study that focuses solely on function inlining.
	We investigate the complex decision-making process 
	for function inlining within the LLVM compiler toolchain, 
	diving into the inner heuristics, compiler options, 
    and	the cost model for inlining.
    \item 
    We clarify the misleading subtleties
    in function inlining practices, which
    have been underestimated
    or overlooked.
    \item
    We present extreme inlining with 
    a different combination of compiler 
    options, and how it can distort 
    static features 
    (\eg machine code, 
    control flow graphs).
    \item 
    We conduct an extensive study on 
    security implications by investigating
    how (extreme) inlining affects 
    ML-assisted binary reversing tasks.
 \end{itemize}

To facilitate future research, 
we release our datasets,
analysis, and models as 
open source~\footnote{\url{https://doi.org/10.5281/zenodo.17759528}}.

\section{Background}
\label{sec_bk}

\begin{table}[t!]
    \centering
    \caption{Special 
    LLVM options to emit intermediate
    outputs of function inlining.}
    \resizebox{0.99\columnwidth}{!}{
    \begin{tabular}{ll}
        \toprule
        
\textbf{Option} & \textbf{Description} \\
\midrule
\cc{-Rpass=inline} & Shows inlining information (\eg success, failure)\\
\cc{-Rpass-analysis=inline} & 
Shows inlining analysis (\eg cost, threshold) \\
\cc{-Rpass-missed=inline} & Shows missed inlining (\eg heuristics, high cost) \\
\bottomrule

    \end{tabular}
    }
    \label{tab:developer_option_inlining}
    \vspace{-15px}
\end{table}

\PP{LLVM Architecture and Intermediate Outputs}
LLVM (Low-level virtual machine)~\cite{llvm14} 
is a collection of compiler toolchains, which
provides modular building blocks 
for both the frontend and backend.
Its core design centers around 
an intermediate representation (IR), which
enables support for 
various programming languages at 
the frontend,
analysis and transformation of IRs 
at the middle end, and
target-specific code generation 
at the backend.
LLVM provides numerous options 
for emitting intermediate
results during transformations~\footnote{GCC offers 
the \cc{-fdump-ipa-inline} option for 
determining function inlining 
in a single-source code~\cite{1to1_1ton}
alone.}. 
In this work, we adopt three special options in LLVM 
(\autoref{tab:developer_option_inlining})
for gathering diagnostic reports on function inlining.
Although the outputs may be imperfect~\cite{rpass_limitations},
we harness them to reveal how 
a cost model internally determines
whether a function can be inlined 
through complex computations (\ie cost, threshold).
As a final note, we explore the LLVM source
for in-depth analysis of its function inlining optimization.

\PP{Symbols and Linkage Types}
The compiler maintains several different linkage 
types that define the visibility of 
a symbol when emitting a binary,
which determine how they can be accessed across different modules. 
These types are crucial to handling symbol visibility, code optimization, 
and symbol collision.
First, a symbol with an external linkage 
is supposed to be visible
outside of a module for being referenced 
by other modules,
including functions and variables without the \cc{static} keyword
or with \cc{extern}.
A function is declared with an external linkage 
by default~\cite{api_design_c++}.
Second, a symbol with an internal linkage 
is only visible within a module,
which cannot be referenced from others (but avoid name collisions).
It is noted that static functions are considered 
as unused global symbols (thus safely removed~\cite{inling_code_size}) 
once they are inlined.
Third, a symbol with a private linkage is 
analogous to internal ones,
with an additional constraint that even 
link time optimization (LTO) cannot make it 
visible to other translation units.
Lastly, a symbol with a weak linkage allows for 
multiple definitions
of the identical symbol across different modules, which the linker chooses from afterward.
This flexibility renders the multiple symbols 
in need present for inline functions 
or template instantiations in C++.

\PP{DWARF Information}
Debugging With Attributed Record 
Formats (or DWARF for short)~\cite{dwarf5}
defines a %
structured data format that contains
essential information for debugging
such as mappings between source and 
machine code.
A debugging information 
entry (DIE) enables one to create a low-level 
representation of a source program,
each of which includes an identifying tag 
and a series of attributes. 
The tag specifies the class of the entry while 
the attributes define its specific 
characteristics. 
\autoref{tab:DIEtagat} in Appendix enumerates the 
tag and attributes 
that are associated with the inlining
behavior of a function in DWARF. 
The information can be generated during 
compilation with the \cc{-g} option. 
As DWARF provides a unified approach 
that is orthogonal to the programming language 
and its underlying structure,
it has been widely adopted to obtain
the ground truth of an executable binary.
Likewise, this work also utilizes 
DWARF v5~\cite{dwarf5}
to obtain the ground truth of function 
inlining (Section~\ref{sec:implem}).

\PP{Benefits and Downsides of Function Inlining} 
A function inlining optimization 
involves trade-offs.
The evident benefit of inlining 
is to reduce the performance overhead
of a function invocation by
eliminating additional stack-relevant 
operations, such as (re)storing  
a value(s) to a register(s) for passing a parameter(s)
and the prologue and epilogue of a function
for adjusting stack and base pointers.
Next, function inlining may increase instruction cache locality 
(\eg an inlined function in a loop).
Besides, function inlining provides 
a chance for further optimization(s).
Meanwhile, locating duplicate codes 
(\ie instructions) 
multiple times inevitably grows the 
size of an executable binary.
Another downside would be likely to increase
the pressure of allocating more registers
due to their longer liveness and 
more loop invariants.
Lastly, a function inlining process
increases compilation time.

\section{\UP{Motivation and Threat Model}}
\PP{Function Inlining in Prior Work}
\label{priovus_work}
In surveying prior work, we discover that 
function inlining has often been 
misrepresented~\cite{InconvenientTruthsofGroundTruth}, underestimated~\cite{binkit, kim2022reverse, fsmell, discovRE}, 
heuristically addressed~\cite{bingo, asm2vec, yakdan2015no, yakdan2016helping} it, or
\UP{paid limited attention}~\cite{cross_arch_bug_search, 
                    discovRE, 
                    Inlining-vulunerbility, 
                    extracting_cross_platform,
                    xue2019machine, 
                    jin2022symlm, 
                    patrick2020probabilistic, 
                    katz2018using, XDA, BYTEWEIGHT}.
For instance, Alves-Foss~\etal~\cite{InconvenientTruthsofGroundTruth} 
misrepresent the implications of inlining 
by stating that ``\textit{if a source code 
function is inlined, it is no longer 
a function,
and an analysis tool should not 
claim it found that function within the binary}'';
however, inlined functions may still appear in the binary.
Binkit~\cite{binkit} and discovRE~\cite{discovRE} 
underestimate the pervasiveness of 
inlining by applying
the \cc{-fno-inline} option to prevent inlining.
Besides, the \cc{always\_inline} directive 
can override this setting. 
Meanwhile, Dispatch~\cite{kim2022reverse} 
examines a case where a large 
trigonometric function (\eg~\texttt{tan}) 
in firmware 
avoids function inlining to allow 
for handy vulnerability patching.
Many existing works, including 
those focused on tasks like 
bug discovery, code search, and 
code similarity detection~\cite{cross_arch_bug_search, 
Inlining-vulunerbility, extracting_cross_platform, 
xue2019machine, jin2022symlm, 
patrick2020probabilistic, katz2018using}
\UP{do not carefully examine the impact} 
of function inlining, 
failing to account for its potential influence on
code semantics.
Similarly, XDA~\cite{XDA} and BYTEWEIGHT~\cite{BYTEWEIGHT} 
rely on function prologues and epilogues,
without thoroughly discussing the impact of inlining.
Dream(++)~\cite{yakdan2015no, yakdan2016helping} 
addresses inlining heuristically by building signature databases 
to recognize (known) commonly inlined
library functions such as
\cc{strcpy}, \cc{strlen} and \cc{strcmp}.
On the other hand, 
CI-Detector~\cite{Cross-Inlining} and 
Jia~\etal~\cite{1to1_1ton} underscore 
handling inlining directives.
Collberg~\etal~\cite{collberg1997taxonomy} 
describe inlining as an effective and 
practical obfuscation technique
because it removes procedural abstraction 
from the program.
Meanwhile, FUNCRE~\cite{ahmed2021learning} 
addresses inlined library 
functions consistently across
various optimization levels 
but encounters challenges
when multiple consecutive library functions 
are inlined together.
\UP{This work focuses on the unique impact of inlining, 
such as its ability to alter call graphs, control-flow 
graphs, and function boundaries.
}

\PP{\UP{Threat Model and Assumptions}}
\UP{
Our threat model assumes that 
a benign, standard build process can be
repurposed for malicious ends. 
Under this regime, attackers need not
rely on obfuscation to evade detection; 
they can simply recompile the same 
code under different optimization strategies
or build configurations. 
We focus on \emph{realistic}, flag-level 
control of inlining 
(\eg optimization levels and inliner parameters) 
without modifying the compiler's internal logic. 
Notably, this assumption differs 
from traditional obfuscation, 
which involves deliberate, nonstandard 
transformations that introduce
different trade-offs and detection surfaces,
and may lead to performance or 
compliance issues. 
}

\section{Decision Pipeline for Function Inlining}
\label{s:decision-pipeline}

\begin{figure*}[t]
    \centering
        \includegraphics[width=0.75\linewidth]{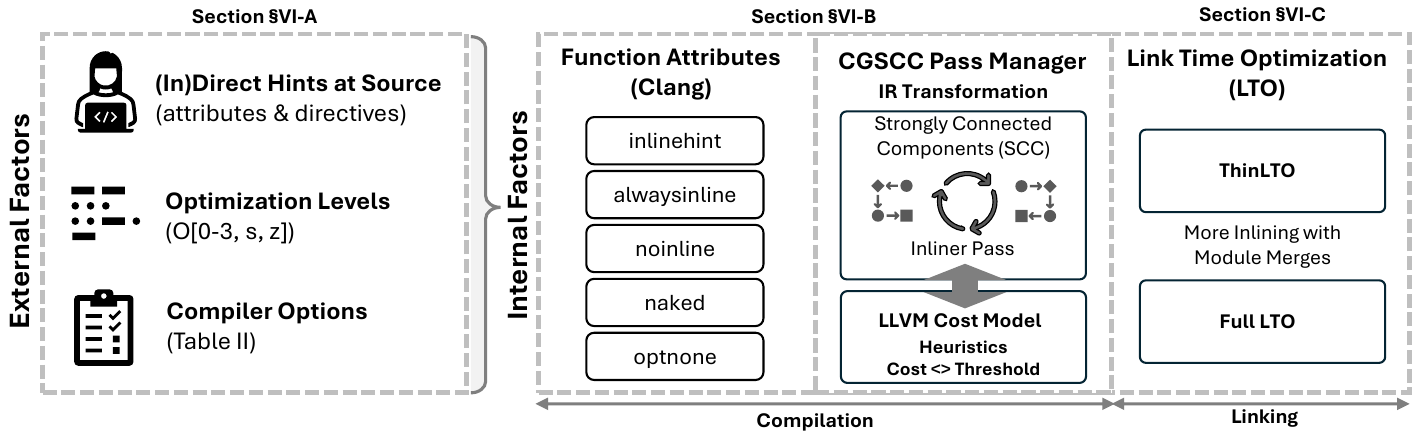}
    \caption{
    Overview of the whole function inlining pipeline in the LLVM toolchain~\cite{llvm14}. 
    We simplify the pipeline of a complex decision process by classifying
    the factors that affect inlining into external and internal elements.
    Varying directives and attributes (\eg \cc{inline} keyword) at source code, optimization levels,
    and compilation options allow a programmer to impact the inlining ratio in a binary (Section~\ref{external_factors}). 
    At compilation time, Clang defines the internal 
    attributes of a function based on those factors.
    Opt iteratively performs function inlining
    across strongly connected components (SCCs) during the transformations of intermediate representation (IR)
    (Section~\ref{internal_factors}).
    Note that the process highly involves with the heuristics and 
    the built-in cost model (\eg checking if a 
    cost is higher than a threshold).
    In the case of link time optimization (LTO) enabled, 
    additional function inlining can be
    performed by consolidating modules
    (Section~\ref{inlining_linking}).
    }
    \label{fig:full_framework}
    \vspace{-10px}
\end{figure*}

\autoref{fig:full_framework} illustrates 
the whole decision pipeline for
function inlining, focusing on
the LLVM~\cite{llvm14} 
compiler infrastructure.

\begin{table*}[t!]
    \centering
    \caption{
    Comprehensive compiler options that affect
    a cost model for function inlining at
    the frontend and middle end.
    }
    \resizebox{0.99\linewidth}{!}{
    \begin{tabular}{llrl}
        \toprule
        \textbf{Option Name} &\textbf{Tool} &\textbf{Default} &\textbf{Description} \\\midrule
\cc{-finline-functions} &Clang &Disabled &Inlines a suitable function based on an optimization level \\
\cc{-finline-hint-functions} &Clang &Disabled &Inlines a function that is (explicitly or implicitly) marked as \cc{inline} \\
\cc{-fno-inline-functions} &Clang &Disabled &Disables function inlining unless a function is declared with \cc{always\_inline} \\
\cc{-fno-inline} &Clang &Disabled &Disables all function inlining except \cc{always\_inline} \\

\midrule

\cc{-inlinedefault-threshold} &Opt &225 &Sets the initial threshold for O1 and O2 alone \\
\cc{-inline-threshold} &Opt &225 & Sets the initial threshold for all optimization levels \\
\cc{-inlinehint-threshold} &Opt &335 &Sets the threshold of a function marked with the \cc{inlinehint} attribute \\
\cc{-inline-cold-callsite-threshold} &Opt &45 &Sets the threshold of a cold call site \\
\cc{-inline-savings-multiplier} &Opt &8 &Sets the multiplier for cycle savings during inlining \\
\cc{-inline-size-allowance} &Opt &100 &Sets the size allowance for inlining without sufficient cycle savings \\
\cc{-inlinecold-threshold} &Opt &45 &Sets the threshold for a cold call site \\
\cc{-hot-callsite-threshold} &Opt &3,000 &Sets the threshold for a hot call site \\
\cc{-locally-hot-callsite-threshold} &Opt &525 &Sets the threshold for a hot call site within a local scope \\
\cc{-cold-callsite-rel-freq} &Opt &2 & Sets a maximum block frequency for a call site to be cold (no profile information) \\
\cc{-hot-callsite-rel-freq} &Opt &60 & Sets a minimum block frequency for a call site to be hot (no profile information) \\
\cc{-inline-call-penalty} &Opt &25 &Sets a call penalty per function invocation \\
\cc{-inline-enable-cost-benefit-analysis} &Opt &false & 
Enables the cost-benefit analysis for the inliner \\
\cc{-inline-cost-full} &Opt &false &Enables to compute the full inline cost when the cost exceeds the threshold \\
\cc{-inline-caller-superset-nobuiltin} &Opt &true & Enables inlining when a caller has a superset of the callee's \cc{nobuiltin} attribute\\
\cc{-disable-gep-const-evaluation} &Opt &false & Disables the evaluation of \cc{GetElementPtr} with constant operands \\
\bottomrule

    \end{tabular}
    }
    \label{tbl:flags}
    \vspace{-10px}
\end{table*}

\subsection{External Inlining Factors}
\label{external_factors}

\subsubsection{(In)Direct Hints on Inlining at Source} 
The C and C++ programming languages 
offer (in)direct means for a programmer
to hint function inlining using
a directive or an attribute.
As part of the C and C++ standards, the \cc{inline} specifier provides 
a hint to inline a function 
at the function declaration~\footnote{The \cc{\_\_inline} specifier serves 
the same purpose before 
the C99 standard; however, it is still 
compatible with modern compilers.}.
Besides, different compilers provide 
additional attributes and pragmas with
respect to inlining.
Clang and GCC allow the compiler to 
explicitly inline the function with the 
specific function attribute of \cc{\_\_attribute\_\_((always\_inline))}
regardless of the optimization level
(\ie \cc{-O0} is not an exception).
Conversely, \cc{\_\_attribute\_\_((noinline))} 
prevents the compiler from inlining the function
~\footnote{Similarly, \cc{Microsoft Visual C++} 
compiler supports the
\cc{\_\_forceinline} and 
\cc{\_\_declspec(noinline)} keywords
to enforce and prevent
function inlining, respectively.
}.
In a similar vein, both \cc{\#pragma inline}
and \cc{\#pragma noinline} directives serve the
purpose of impacting function inlining.
Note that the above compiler-specific directives
are not part of the standards, which may differ
depending on their implementations.
\sloppy Akin to direct hints above, 
other attributes can indirectly 
influence function inlining. 
The \cc{\_\_attribute\_\_((flatten))} 
attribute allows for inlining all 
callees within a function as if
each call site had 
\cc{\_\_attribute\_\_((always\_inline))}. 
Meanwhile, the 
\cc{\_\_attribute\_\_((naked))} attribute 
is typically used for an assembly function, 
indirectly causing the compiler to avoid 
function inlining. 
Similarly, the 
\cc{\_\_attribute\_\_((optnone))} 
attribute disables all optimizations 
for a function as well as inlining.

\subsubsection{Optimization Levels}
A compiler optimization level 
highly affects IR and machine code
generation, striking a balance 
between performance, compilation time,
and size. 
The \cc{-O[0-3]} typically represents
the level of optimization (\ie none,
basic, medium, or maximum) whereas 
the \cc{-O[s,z]} 
adjusts the code size (\ie small or minimum).
The optimization strategy unavoidably 
impacts on function inlining:
\eg~\cc{-O3} performs aggressive inlining to
enhance execution speed, \cc{-Os} does moderate
inlining to balance performance and size, 
and \cc{-Oz} attempts to minimize code size
(\eg less inlining than others).

\subsubsection{Compiler Options}
LLVM~\cite{llvm14} offers varying options
that configure global settings for 
inlining decisions across all functions.
\autoref{tbl:flags} summarizes
comprehensive (style-agnostic) options that 
assist in elaborately controlling
the function inlining behaviors in LLVM
~\footnote{The supportive options can be found with 
\cc{--help-hidden}. 
The function-inlining-specific options 
available for the LLVM optimizer (opt) 
can be passed 
with \cc{-mllvm [flagname=value]} in Clang. 
Note that we exclude
the options that rely on a specific
style or language (\eg GNU89, Assembly).}.
For instance, \cc{-finline-functions}
allows the compiler to decide on a function
to be inlined.
Alternatively, \cc{-finline-hint-functions} 
instructs the compiler to consider a 
function with a specific keyword or 
attribute for inlining alone. 
In contrast, Clang also provides the options 
to disable inlining based on compiler heuristics 
with \cc{-fno-inline-functions} or to disable 
it entirely, except for a function 
marked \cc{alwaysinline}
(Section~\ref{internal_factors}).

\subsection{Internal Inlining Factors at Compilation}
\label{internal_factors}

\subsubsection{Function Attributes}
Based on the external factors like hints, optimization levels, and compiler options, 
Clang internally labels 
the attribute of a function as one of the following: 
\cc{inlinehint}, \cc{alwaysinline}, 
\cc{noinline}, \cc{naked}, and \cc{optnone}.
Setting those attributes
occurs at the early stage of IR
generation (from AST), conforming
custom directives in a subsequent
optimization process.
For instance, although \cc{-O0} disables most optimizations for compilation speed and debugging,
it performs minimal and necessary transformations (\eg \cc{alwaysinline})
to ensure correct code execution. 
As a final note, 
a compiler-generated function like 
an intrinsic function is often marked with
\cc{alwaysinline} to ensure efficiency.

\subsubsection{LLVM Inliner Pass}
LLVM offers the \emph{inliner pass} 
that can analyze 
and transform intermediate representations, 
which internally maintains a pass manager 
for applying a chain of passes to IRs 
in a specific order~\cite{LLVMPasses}. 
The LLVM's function inliner pass performs 
the decision of function inlining, 
which is tightly coupled with the 
CGSCC pass manager~\cite{CGSCCPassManager} that runs on a strongly connected 
component (SCC) in a call graph.
The SCC consists of callers and callees, where each function is simplified via
a sequence of transformation passes such as control flow graph simplification (SimplifyCFG pass~\cite{SimplifyCFGPass}), 
scalar replacement of aggregates (SROA pass~\cite{SROAPass}), and early common sub-expression elimination (EarlyCSE pass~\cite{EarlyCSEPass}).
These preliminary steps ensure that non-trivial 
SCCs are adequately optimized 
prior to function inlining. 
Finally, CGSCC iteratively invokes
the function inliner pass across SCCs
in a bottom-up order based on a cost model
(See Section~\ref{sss:cost-model}).

\begin{table}[!t]
\centering
\caption{Selective call-site cases that
never inline a function as 
the built-in heuristics in LLVM. 
}
\resizebox{\columnwidth}{!}{
    \begin{tabular}{p{0.15\columnwidth}p{0.85\columnwidth}} 
    \toprule
            \textbf{Scope} & \textbf{Case Description}      \\ 
\midrule
Basic Block       &  A misuse of a block address outside of the specific instruction (\eg \cc{callbr}) \\ \midrule
Caller            & 
\cc{optnone} attribute that disables all optimizations including inlining             
\\ 
 & \cc{noduplicate} attribute that avoids multiple instances
 \\ \midrule
Caller and Callee
& Both callee and caller with conflicting attributes (\eg \cc{alwaysinline} and \cc{noinline} together)                                                                               \\ 
 & An incompatible null pointer definition between the caller and callee                                                                                                                                           \\ \midrule
Callee  & 
\cc{noinline} attribute                                                                                                                               \\ 
            & An interposable function that can be replaced or overridden by another at link time or runtime                                                                                                       \\ 
            & A unsplit coroutine call that cannot split into separate parts for suspension and resumption %
            \\ \midrule
Call site         &  
\cc{noinline} attribute
\\ 
        & A \cc{Byval} argument with an incorrect memory address space (\ie without the \cc{alloca} address space)                                                                                                                       \\ 
       & An indirect call where the function being called cannot be determined at compile time                                                                                                                                  \\ \midrule
Instruction       &  Amount of memory unknown at compile time (\eg dynamic allocation)                                                                                                                                                      \\ 
       & A function with multiple return paths (returns twice)                                                                                                                                                             \\ 
       & A function initialized with variadic arguments                                                                                                                                                            \\ 
       & An indirect branch where the target is determined at runtime                                                                                                                               \\ 
       & An intrinsic that is too complex to be inlined (\eg \cc{icall\_branch\_funnel} or \cc{localescape}).                                                                                                                              \\ 
       & A recursive call that lead to infinite loops or allocate too much stack space                                                                                                                         \\ 
\bottomrule

    \end{tabular}
}
\label{llvm_no_inlining}
\vspace{-10px}
\end{table}

\subsubsection{Cost Model}
\label{sss:cost-model}
\begin{table}[t!]
    \centering
    \caption{Conditions that affect a cost 
    and a threshold for function inlining
    for the LLVM's cost model.
    }
    \resizebox{0.99\linewidth}{!}{
    \begin{tabular}{lcl}
        \toprule

\textbf{Target} & \textbf{Impact} & \textbf{Condition} \\ 
\midrule

\multirow{9}{*}{Cost} &

\multirow{6}{*}{$\uparrow$} & Missing instruction simplification (\eg switch, loop, load) \\
& & Presence of an intrinsic function\\
& & Unoptimized call sites within a callee \\
& & Complex memory operations that cannot be simplified \\
& & Cold calling conventions (\cc{coldcc}) \\
& & General function call overhead (\cc{call penalty}) \\
\cmidrule(lr){2-3}
& \multirow{3}{*}{$\downarrow$}
& Last call of a static function \\
& & Call site arguments with values (\eg \cc{byval}) \\
& & Successful transformation
(\eg indirect to direct calls)  \\

\midrule
\multirow{7}{*}{Threshold} & 
\multirow{3}{*}{$\uparrow$} & Presence of the
\cc{inlinehint} attribute \\ 
 & & Hot region \\ 
 & & Adjustments based on the target-specific architecture \\ 
\cmidrule(lr){2-3}
& \multirow{4}{*}{$\downarrow$} 
&  Presence of the \cc{minsize} or \cc{optsize} attributes  \\
&  & Cold region  \\ 
&  & Complex branching \\ 
&  & Presence of vectorization instructions \\ 
\bottomrule

    \end{tabular}
    }
    \label{tbl:cost_threshold}
\end{table}
The cost model in LLVM begins with 
the pre-defined cases
that must not be inlined (as heuristics).
\autoref{llvm_no_inlining}
enumerates such cases with a scope.
Once the model confirms feasible cases 
for function inlining, the function 
inliner pass evaluates a call site
(\ie a pair of caller and callee) to 
dynamically compute 
a \emph{threshold} and a \emph{cost}
depending on the aforementioned factors.
Simply put, the inliner pass
mechanically performs
inlining transformation 
in SCC when \emph{the cost is less
than the threshold}.

\begin{figure}[t!]
    \centering
    \resizebox{0.9\linewidth}{!}{%
        \includegraphics[width=\linewidth]{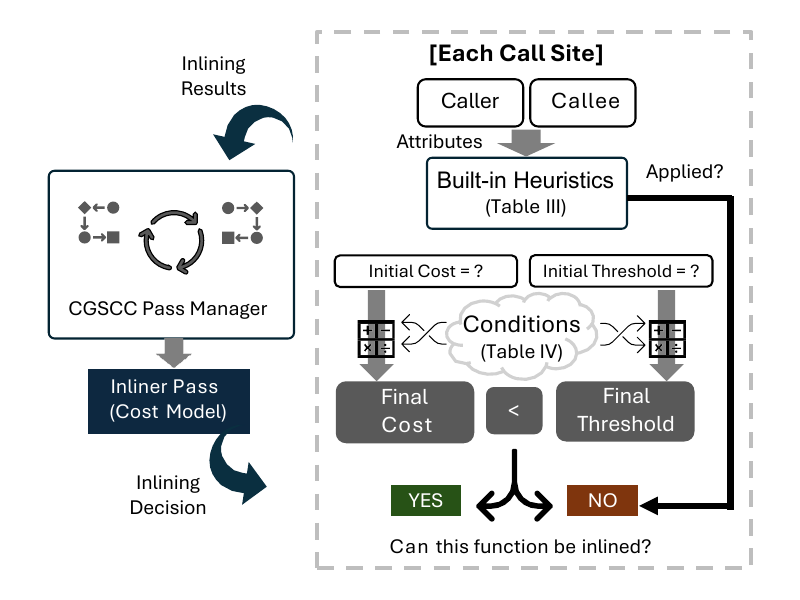}
    }
    \caption{
    Overview of the cost model in LLVM
    for determining a function to be inlined.
    While the CGSCC pass manager runs on 
    strongly connected components, 
    it invokes the inliner pass when needed.
    The cost model first considers the heuristics (\autoref{llvm_no_inlining})
    of each call site to check if it can never be inlined.
    If possible, a cost and a threshold are initialized,
    followed by being updated according to
    dynamic conditions (\autoref{tbl:cost_threshold}).
    The process mechanically determines 
    function inlining according to a cost.
    Then, CGSCC iteratively performs further
    optimizations based on the inlining.
    }
    \label{fig:cost_model}
     \vspace{-10px}
\end{figure}

\PP{Threshold Computation}
At each call site, a threshold is initialized by an optimization 
level (\eg \cc{-Oz} $\rightarrow 5$, \cc{-Os} $\rightarrow 50$, \cc{-O1} $\rightarrow 225$, \cc{-O2} $\rightarrow 225$, \cc{-O3} $\rightarrow 250$).
Additionally, the CGSCC pass manager holds fruitful information
on the call site with profile summary information 
(PSI; \eg hotness or coldness of a function, execution times)
and block frequency information (BFI; \eg execution frequency of
an individual basic block).
Afterward, the inliner pass increases
or decreases the threshold
depending on \WC{1} the attributes of a caller and/or a callee,
\WC{2} the property of a code region, and \WC{3}
other constructs like a complex branch or a vectorized instruction
(\autoref{tbl:cost_threshold}).
For instance, the \cc{inlinehint} attribute via the \cc{inline} specifier 
can increase the threshold.

\PP{Cost Computation}
By default, a cost is initialized to zero. 
A special case arises when a callee is a 
static function and the call site is the
last call invocation, setting the initial cost
to a negative value of $-15,000$.
This substantially increases 
the likelihood of function inlining.
As summarized at \autoref{tbl:cost_threshold}, various conditions can adjust the cost.
In essence, the inliner pass walks through
each basic block and every instruction by adding
or subtracting a designated value, obtaining
the final cost.
It is noteworthy to mention that a cost or
a threshold could be configured by 
compiler options (\autoref{tbl:flags}).
On the one hand, the cost can be penalized 
(\ie increasing) in cases of missing 
instruction simplification, presence of 
an intrinsic function, an optimized
call site, a complex memory operation, or a
cold calling convention.
On the other hand, the cost can 
decrease in cases of call site arguments
with values and transformation from
indirect to direct call invocation.
LTO promotes every symbol to internal linkage 
using the \cc{internalize} pass, attempting to eliminate it.
However, the inlined function 
under LTO may remain 
if it has not been inlined 
at all call sites.

\subsection{Internal Inlining Factors at Linking}
\label{inlining_linking} 

\subsubsection{Link Time Optimization and Inlining}
Enabling LTO~\cite{gcc_lto}
literally provides the linker with global 
optimization opportunities from the whole program viewpoint at link time,
which otherwise cannot be performed under the individual compilation unit.
There are two primary types of LTO~\footnote{To enable 
LTO in LLVM, we use the \cc{-flto} option using the 
LLVM linker (\ie \cc{-fuse-ld=lld}).}. 
in LLVM: ThinLTO~\cite{ThinLTO} and Full LTO~\cite{full_lto}, 
each with different implications for function inlining.

\PP{Inlining with Full LTO} 
Full LTO~\cite{full_lto} 
enables aggressive inlining 
across the entire program by making 
all functions globally visible, allowing for 
further optimizations such as 
constant propagation and dead code elimination. 
However, it incurs significant 
resource overheads (\eg memory usage), 
taking a longer compilation time.

\PP{Inlining with ThinLTO} 
ThinLTO~\cite{ThinLTO} aims to reduce computational 
overheads while adopting the benefits 
of Full LTO.
In essence, ThinLTO generates a summary of the functions in each module,
which allows for performing additional function inlining 
along with incremental cross-module information during the linking phase.

\section{Subtleties in Function Inlining Practices}

In this section, we clarify 
six (common) misleading beliefs about
function inlining, none of
which hold true in practice.
Then, we introduce the extreme 
inlining strategy.

\subsection{Misleading Case Study}

\PP{A function would not be inlined with 
\cc{-O0} or the \cc{-fno-inline} option}
One may want to generate a 
binary without any optimization
(including inlining), thereby 
preserving a function and its inner structure
eases debugging (\eg tracing an execution flow,
identifying an error~\cite{no-inline-functions-flag}).
Although the compiler conforms with the request
to suppress most optimizations, a function
with the \cc{always\_inline} directive remains
an exception regardless of a given optimization
level or provided compiler option.
In a nutshell, both \cc{-O0} and the \cc{-fno-inline} option 
cannot override the above directive, 
while it broadly prevents inlining 
as in \autoref{fig:binaries_cdf} and \autoref{fig:coreutils_clang_flags}.

\PP{The \cc{always\_inline} directive would always inline a function}
Any function directive (including \cc{inline} and \cc{always\_inline})
does not necessarily guarantee a function to be inlined because
the compiler will make a final decision 
throughout the decision pipeline 
(\autoref{fig:full_framework})
to evaluate a call site with the complex cost model (\autoref{fig:cost_model}).
There are plenty of such cases 
where the compiler never considers function inlining
irrespective of optimization levels, directive hints, 
or compiler options,
including recursive calls, indirect calls, 
and functions with variadic arguments
(\autoref{llvm_no_inlining}). 

\PP{A function without any inlining-relevant
directive would not be inlined} 
In practice, LLVM regards every function 
as a candidate 
to be inlined during compilation, 
considering a function size, complexity,
optimization level, and compiler options as a whole.
In essence, function inlining is an opt-out operation 
in that a function may be inlined 
unless it has an explicit
mark with the \cc{no\_inline} attribute
\autoref{fig:packages} and \autoref{fig:sankey_coreutils} show such examples.

\PP{An inlined function would disappear in a binary}
One of the most common misconceptions is that 
an inlined function has been eliminated 
(hence fully disappeared) because
the call site is replaced with the function.
The answer is partially affirmative; however,
it is possible that an inlined function could 
be present (with a symbol) in a final binary
when the function needs to be invoked globally
\autoref{fig:packages} and \autoref{fig:sankey_coreutils} show such instances.

\PP{A conflicting condition for function inlining
might lead to unpredictable behaviors}
One may pose a hypothetical question like \WC{1} what if the two functions may
have a contradictory condition (\eg one with \cc{inline}
and another with \cc{noinline})? or \WC{2} what if a conflicting option is given
(\eg \cc{-fno-inline}) when compiling a function with \cc{inlinehint})?
To address such issues, the compiler 
establishes several pre-defined rules.
First, the attributes (directives) come over 
the compiler options.
Next, the compiler prioritizes each directive
in the following order: \cc{optnone}, \cc{noinline}, 
\cc{minsize}, \cc{optsize}, \cc{inlinehint}, and 
\cc{always\_inline}.
To exemplify, when a function is declared with 
\cc{inlinehint} or \cc{always\_inline},
\cc{optnone} takes precedence, effectively negating other attributes. 
Similarly, the presence of \cc{noinline} overrides \cc{alwaysinline}.

\PP{A function in a library would not be inlined}
It is common practice to resolve a function 
address in a library at runtime, internally following 
the routine via Procedure Linkage Table (PLT) 
and Global Offset Table (GOT),
with the assistance of a dynamic loader.
However, there are a handful of scenarios that
a library function can be an inlining candidate. 
First, the preprocessor prepares the source
by handing various directives 
(\eg macro definition, file inclusion, conditional 
statement), possibly rendering a function defined in
a header file inlined when the header is included in
multiple translation units.
Second, similarly, it is possible for the compiler to
inline a function if a header-only library~\cite{header_Function_Inlining, inline_c_rules} provides
complete function definitions visible in a header.
Third, LLVM defines a series of special 
intrinsic functions~\cite{intrinsic_func} (\eg \cc{llvm.memcpy}, \cc{llvm.memset})
that operate on the IR level, being exposed as
an inlining candidate with the \cc{always\_inline} 
attribute~\footnote{The intrinsic functions in LLVM
are beneficial, providing code simplification and 
platform independence as well as leveraging the features 
of a target architecture.}.

\subsection{Inlined but Remaining Functions}
A function symbol with an internal linkage 
(\eg \cc{static})
goes away when inlined in every call site.
However, even with an internal function, 
the function may remain after
inlining (from elsewhere) if 
any call site requires 
that function invocation.
Conversely, a function with an 
external linkage (\eg \cc{extern})
would remain after being inlined by default.

\subsection{Extreme Inlining}
Distinguishing from aggressive inlining by high optimization levels, 
we introduce the concept of
\emph{extreme inlining}
that promote the
function inlining ratio 
beyond standard compiler behaviors.
Section~\ref{ss:lto_fn_inl} elaborates 
our exploration by exploiting 
a decision pipeline
to establish 
a strategy for extreme inlining, including
opt-levels, compilation options, and LTO.

\section{Function Inlining and Security Implications}
\label{s:eval}

We run our experiments on a 64-bit Ubuntu 20.04 system equipped with 
Intel(R) Xeon(R) Gold 5218R CPU 3.00GHz, 512GB RAM, 
and two RTX A6000 GPUs. 

\PP{Benign Software Corpus}
The first half of 
\autoref{tab:binary_dataset}
summarizes our benign program corpus 
to evaluate the impact of
function inlining on ML-based
models.
Each package and application
may have been compiled with slightly 
different compilation options 
(\eg \cc{-Oz}, \cc{-O3}) or 
inlining thresholds 
(\eg $2{,}225$, $200{,}000$) 
due to build constraints. 
In cases where LTO
(either thin or full) fails, 
we disable it but preserve 
the original inlining thresholds 
and optimization levels.
To induce extreme inlining behavior, 
we empirically tune the relevant 
compilation settings.
To minimize evaluation bias, 
we \UP{de-duplicate identical functions 
by symbol names and ensure test samples 
never appear in training.}
Our final benign application dataset consists of
$1,524$ program variations,
including additional $398$ samples with extreme inlining applied.

\PP{Malicious Software Corpus}
The second half of \autoref{tab:binary_dataset} 
summarizes our malware corpus.
We focus on IoT malware due to its prevalence, 
shared code bases~\cite{genealogy_malware}, 
and the availability of leaked source code 
(\eg Mirai~\cite{mirai_botnet} and Gafgyt~\cite{gafgyt}). 
For detecting malware,  
we construct a curated dataset of 
$761$ malware samples from 
VirusShare~\cite{virusshare}, including 
Mirai, Gafgyt, and
Tsunami, by removing corrupted binaries, and 
packed executables from a broader collection.
Similarly, 
for classifying malware,
we prepare
$12,727$ samples across 10 families from Alrawi \etal~\cite{alrwai_circle}: 
Mirai, Gafgyt, Tsunami, Lotoor, 
Dofloo, DDoSTF, ExploitScan, Dvmap, 
Gluper, and Healerbot. 
To evaluate the impact 
(\ie evading detection) 
of function inlining, 
we recompile open-sourced 
Mirai and Gafgyt
($100$ samples per each)
by applying extreme inlining 
with custom compiler options.
Our final malware dataset 
consists of $13,688$ samples
in total.
\UP{For malware-related tasks,
we remove duplicate samples
that are identical 62-dimensional 
statistical feature representation
to avoid model overfitting.
As a final note, \autoref{tab_extreme_inlining_recipe}
in Appendix shows the whole recipe for extreme inlining
in our experiments.
}

\subsection{Security Impact of Function Inlining 
on ML Models}
\label{ss:eval2}

\PP{Research Questions}
We define five research questions 
to evaluate how function inlining 
affects ML-based 
models from a security aspect.
We choose five ML-based security tasks, including
binary code similarity detection (T1), 
function name prediction (T2),
malware detection (T3),
malware family classification (T4), 
and vulnerability detection (T5).
\begin{itemize}[leftmargin=*]
    \item
    \textbf{RQ1}: How does function inlining affect 
    ML-based binary code similarity detection models (Section~\ref{ss:bcsd})?
    \item 
    \textbf{RQ2}: How does function inlining affect ML-based function symbol name prediction models (Section~\ref{ss:fnp})?
    \item 
    \textbf{RQ3}: How does function inlining affect ML-based malware detection models (Section~\ref{ss:md})? 
    \item 
    \textbf{RQ4}: How does function inlining affect 
    ML-based malware family classification models (Section~\ref{ss:mfp})?
    \item
    \textbf{RQ5}: How does function inlining affect an ML-based vulnerability detection model (Section~\ref{ss:vd})?
\end{itemize}

\begin{table}[t!]
    \centering
    \caption{
    Binary corpus for experiments to assess
    the impact of function inlining on ML-based
    models.
    We choose various security tasks:
    (T1) binary code similarity detection, 
    (T2) function name prediction, 
    (T3) malware detection, 
    (T4) malware family prediction, and 
    (T5) vulnerability detection.
    We adopt two open-sourced malware
    for artificial inlining.
    Note that we generate 32-bit ARM binaries 
    for T4(*), while the others are 
    generated as 64-bit Intel x86 binaries.
    }
    \resizebox{0.99\columnwidth}{!}{
    \begin{tabular}{lllrrrrrrr}
        \toprule

\multirow{2}{*}{\textbf{Type}} &\textbf{Package or} &\multirow{2}{*}{\textbf{Version}} &\textbf{Baseline} &\multirow{2}{*}{\textbf{T1}} &\multirow{2}{*}{\textbf{T2}} &\multirow{2}{*}{\textbf{T3}} &\multirow{2}{*}{\textbf{T4*}} &\multirow{2}{*}{\textbf{T5}} \\
 &\textbf{Application(s)} & &\textbf{Binaries} & & & & &  \\\midrule
\multirow{13}{*}{\textbf{Benign}} &\textbf{coreutils~\cite{coreutils}} &9.3 &106 &742 &742 & 624 & - & - \\
&\textbf{binutils~\cite{binutils}} &2.40 &22 &118 &118 & 78& - & - \\
&\textbf{diffutils~\cite{diffutils}} &3.8 &4 &28 &28 &24 & - & -  \\
&\textbf{findutils~\cite{findutils}} &4.9 &4 &28 &28 &24 & - & - \\ 
\cmidrule(lr){2-9}
&\multirow{2}{*}{\textbf{openssl}~\cite{openssl}} &3.1.4 &1 &7 &7 &6 & - & - 
\\ 
& &1.0.2d &1 & - & - & - & - &6 \\
\cmidrule(lr){2-9}
&\textbf{lvm2~\cite{lvm2_tags}} &2.03.21 &52 &364 &364 & - & - & - \\
&\textbf{gsl~\cite{gsl}} &2.7.0 &2 &14 &14 & 9 & - & - \\
&\textbf{valgrind~\cite{valgrind}} &3.21.0 &4 &28 &28 & 156 & - & - \\
&\textbf{openmpi~\cite{openmpi}} &4.1.5 &7 &49 &49 & 36& - & - \\
&\textbf{putty~\cite{putty}} &0.79 &6 &42 &42 &36 & - & - \\
&\textbf{nginx~\cite{nginx}} &1.21.6 &1 &7 &7 & 6& - & - \\
&\textbf{lighttpd~\cite{lighttpd}} &2.0.0 &2 &14 &14 &10 & - & - \\
&\textbf{SPEC2006 ~\cite{spec_cpu2006}}& - & 15& - & - & 83& - & - \\
\midrule
\multirow{10}{*}{\textbf{Malware}} &\textbf{Mirai~\cite{mirai-source-code}} & - & 2 & - & - &  261 & 9,215 & - \\
& \textbf{Gafgyt~\cite{lizkebab-source-code}} & - & 1 & - & - & 421 & 3,119 & - \\
& \textbf{Tsunami} & - & - & - & - & 79 & 228 & - \\
& \textbf{Lotoor} & - & - & - & - & - & 80 & - \\
& \textbf{Dofloo} & - & - & - & - & - & 40 & - \\
& \textbf{DDoSTF} & - & - & - & - & - & 15 & - \\
& \textbf{ExploitScan} & - & - & - & - & - & 11 & - \\
& \textbf{Dvmap} & - & - & - & - & - & 8 & - \\
& \textbf{Gluper} & - & - & - & - & - & 6 & - \\
& \textbf{Healerbot} & - & - & - & - & - & 5 & - \\
\bottomrule
 
    \end{tabular}
    }
    \label{tab:binary_dataset}
    \vspace{-10px}
\end{table}

\PP{ML Model Selection for Security Tasks}
We select a diverse set of 
ML-based models 
spanning both discriminative and 
generative tasks, 
covering a range of architectures 
(\eg embedding-based models, GNNs, 
Transformers, RNNs) and 
input modalities (\eg statistical
features, dynamic traces).
Note that we use the original 
pre-trained form for T1 and T5.
Meanwhile, we retrained an
\UP{AsmDepictor} model for T2
to match our Clang-based corpus,
as their originals were trained 
on GCC binaries.
\UP{Similarly, SymLM was retrained on
our dataset so that execution-aware embeddings
can reflect the semantics of 
our compilation settings.}
For T3 and T4, we reimplemented 
the model architectures based on 
the descriptions in the original paper~\cite{dl_fhmc},
and retrained the models on our corpus.

\subsubsection{(RQ1) Inlining Impact on BCSD Models}
\label{ss:bcsd}

\begin{figure*}[!htbp]
    \centering
    \resizebox{
    0.99\linewidth}{!}{%
    \includegraphics[width=\linewidth]{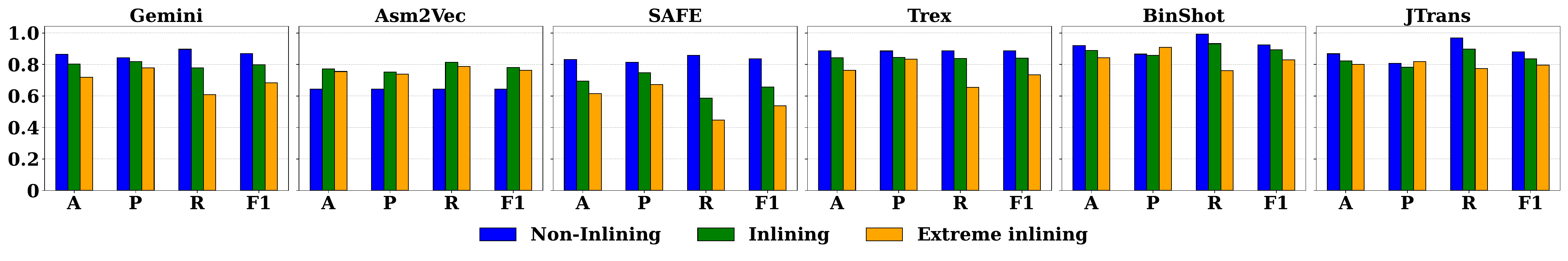}
    }
   
    \caption{
    Experimental results across 
    \UP{six} BCSD models on three datasets:
    non-inlining, inlining, and
    extreme inlining pair samples. 
    We carefully prepare a dataset comprising 
    non-inlining, inlining, and extreme inlining cases 
    to investigate the effectiveness
    of inlining. 
    Overall, there is a gradual decrease 
    in performance for the inlining samples.
    We discuss the exceptional case of Asm2Vec
    in Section~\ref{ss:bcsd}.
    In general, the other models demonstrate a 
    significant drop in recall compared to precision (\ie increasing false negatives).
    A, P, R, and F1
    denote an accuracy, precision, 
    recall, and F1 value, respectively.
    }
    \label{fig:exp_binsim_task}
    \vspace{-10px}
\end{figure*}

A BCSD task determines code similarity 
between a given pair of code snippets.
To \UP{examine the function inlining
impact on performance},
we prepare three different pairs:
non-inlining samples with the pairs of both non-inlining functions,
inlining samples with the pairs of a non-inlining and an inlining function, and
extreme inlining samples with the pairs of a non-inlining and an inlining function 
that applies our extreme inlining strategy 
We use the functions that
appear in every optimization,
highlighting the effectiveness of inlining on various BCSD models.
We choose \UP{six} representative BCSD models 
for evaluation:
Asm2Vec~\cite{asm2vec}, BinShot~\cite{binshot}, 
Gemini~\cite{xu2017neural}, 
Trex~\cite{pei2020trex}, 
\UP{JTrans~\cite{wang2022jtrans}},
and SAFE~\cite{safe}. 
Finally, we prepare $14,984$, \UP{$95,774$}, 
and $23,378$ 
function pairs from BinShot~\cite{binshot}, 
\UP{JTrans~\cite{wang2022jtrans}},
and the BCSD 
benchmark~\cite{marcelli2022machine},
respectively.

\begin{figure}[t!]
    \centering
    \resizebox{
    0.9\linewidth}{!}{%
    \includegraphics[width=\linewidth]{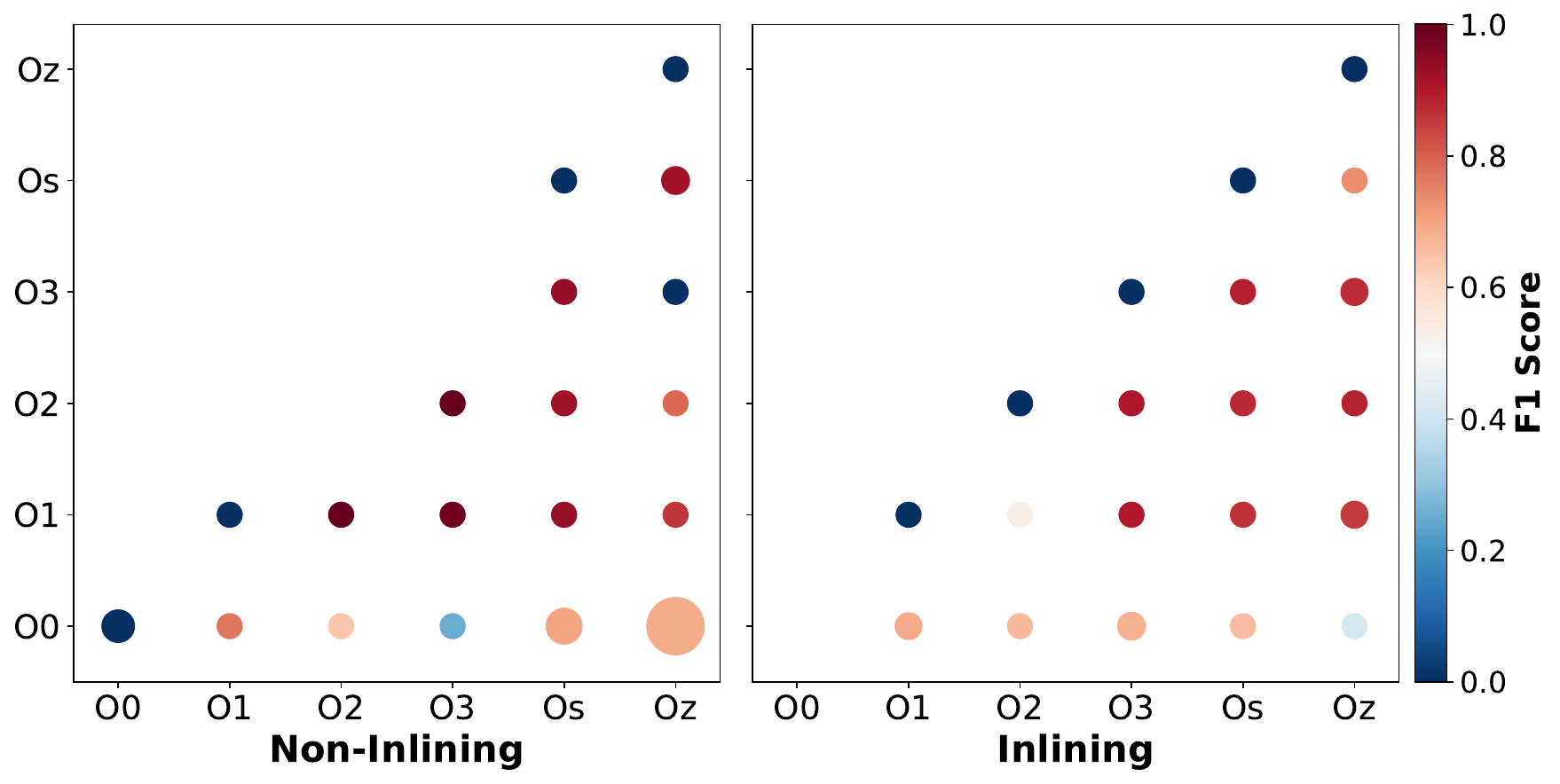}
    }
   
    \caption{
 \UP{
    Asm2Vec performance for the non-inlining (left) 
    and inlining (right) pairs. 
    Circle area indicates the number of pairs,
    and color intensity represents their F1 score.
    The breakdown shows that most
    performance degradation stems 
    from non-inlining pairs
    involving \cc{-O0} versus other optimization levels, 
    suggesting that random walk sequences
    at \cc{-O0} differ substantially from those 
    at higher optimization levels.
    Empty cells correspond to combinations 
    with too few samples.
}
    }
    \label{fig:asm2vec_expl}
    \vspace{-10px}
\end{figure}

\PP{Results}
\autoref{fig:exp_binsim_task} presents 
the performance comparison for \UP{six} 
BCSD models with pairs of
non-inlining, inlining, extreme inlining.
We adopt the BCSD benchmark~\cite{marcelli2022machine} 
that includes Asm2Vec~\cite{asm2vec}, 
Gemini~\cite{xu2017neural}, Trex~\cite{pei2020trex}, and
SAFE~\cite{safe}, providing
a similarity score ($S$) between pairs
in a single platform.
We measure the accuracy, precision, recall, 
and F1 with a threshold of $T=0.5$
(\eg similar if $S > 0.5$)
in comparison with BinShot~\cite{binshot}. 
\UP{In general, moderate performance degradation 
has been observed
(3.6\% in F1; 6.2\% of recall on average) 
for the inlining samples while 
significant decline 
(12.6\% in F1; 21.2\% in recall on average) 
for the samples with extreme inlining.}
Particularly, we observe notable 
decrease in recall (false negatives)
rather than precision (false positives). 
While this suggests that similar 
function pairs may appear dissimilar 
when evaluating inlined functions, 
one could argue that the additional 
semantic information may help 
the model better recognize the 
similarity between functions.
However, as discussed 
in Section~\ref{s:decision-pipeline}, 
function inlining introduces 
opportunities for further optimizations, 
potentially obscuring the similarity
between otherwise a similar function 
pair during evaluation.
Exceptionally, Asm2Vec deviates from 
other models:
\ie non-inlining pair samples exhibits
lower performance than inlining ones.
\UP{
Our investigation reveals that
pair-wise performance discrepancies
arises from the feature
of statically tracing 
(\ie potential execution paths) by 
random walk to infer code similarity
as illustrated in \autoref{fig:asm2vec_expl}.
We observe that the pairs of 
\cc{-O0} versus \cc{-Oz} 
that account for almost half 
(10,614 out of 23,378)
lead to overall performance degradation.
We attribute this effect to differences 
in the random-walk sequences 
generated under \cc{-O0} compared to 
those under higher optimization levels.
}

\subsubsection{(RQ2) Inlining Impact on Function Name Prediction Models} 
\label{ss:fnp}
\begin{figure}[t]
    \centering
    \resizebox{0.99\linewidth}{!}{%
    \includegraphics[width=\linewidth]{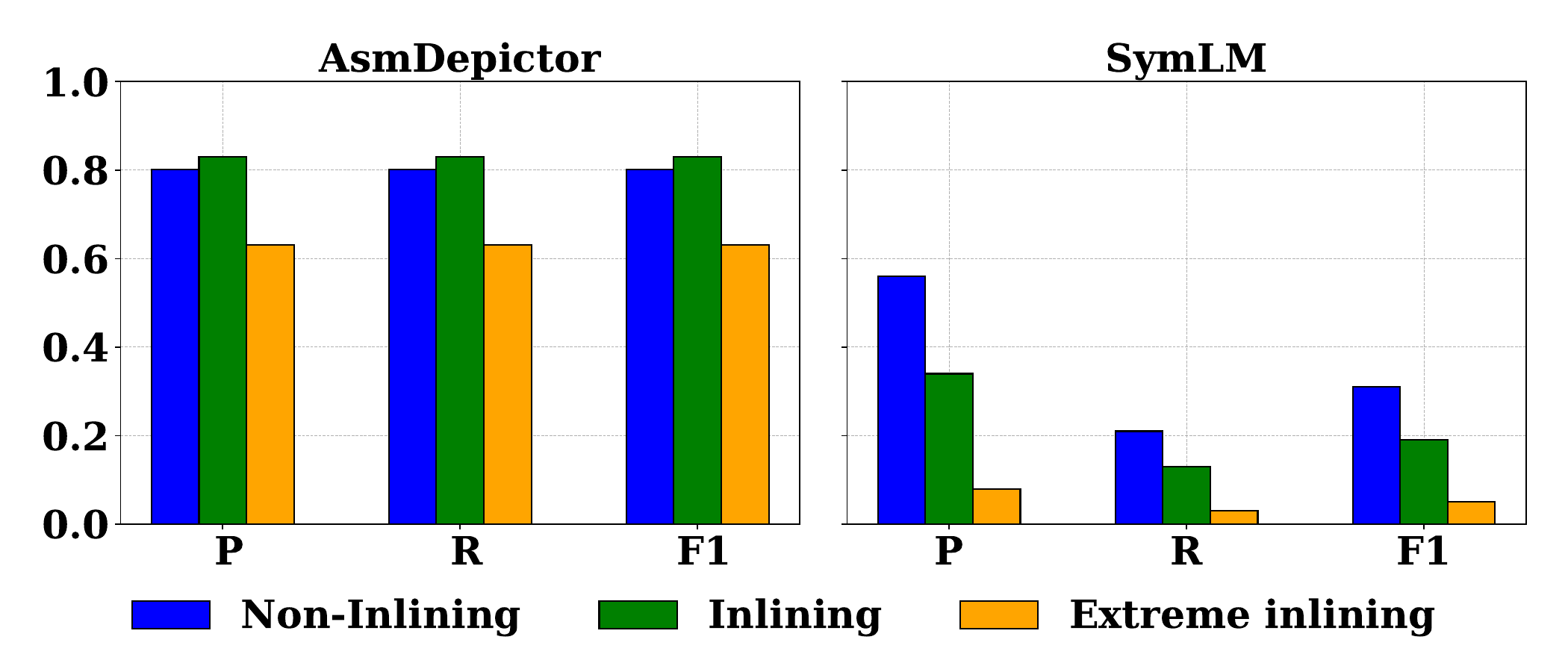} %
    }
    \caption{
    Experimental results with 
    AsmDepictor~\cite{asmdepictor} \UP{and 
    SymLM ~\cite{jin2022symlm}} on three datasets: non-inlining, inlining, and extreme inlining (Section~\ref{ss:fnp}).
    \UP{
    AsmDepictor shows a slight performance 
    increase with the inlining cases, 
    demonstrating that it can leverage the 
    additional contextual information for 
    inference.
    SymLM demonstrates a decrease in performance for the inlining case.
    Meanwhile, the extreme inlining cases 
    mislead both models with unseen patterns, 
    resulting in non-negligible performance 
    drops (\eg 23.99\% for AsmDepictor, 
    75.69\% for SymLM).
    }
    }
    \label{fig:exp_asmdepictor}
    \vspace{-10px}
\end{figure}
A function name prediction task 
aims to infer the original symbol name
of a function given a chunk of assembly code.
Similar to the BCSD experiments, 
we prepare our dataset 
(\autoref{tab:binary_dataset}) for
function name inference.
We choose \UP{two representative models: AsmDepictor~\cite{asmdepictor} (generative) and 
SymLM~\cite{jin2022symlm} (discriminative)}.
We use precision \UP{(P)}, recall \UP{(R)}, 
and F1 for evaluation.

\PP{Results}
\UP{
\autoref{fig:exp_asmdepictor}
presents the empirical results for 
AsmDepictor~\cite{asmdepictor} and 
SymLM~\cite{jin2022symlm} on
our test dataset (\eg 7,863 cases).
AsmDepictor demonstrates improved performance with
inlining, effectively leveraging the additional contextual
information provided.
While curious readers may attribute
AsmDepictor's improved inlining performance
to training data distribution where
79\% of our training dataset consists of
non-inlining functions.
However, both AsmDepictor and SymLM experience
significant performance degradation under
extreme inlining: \eg 24.0\% and 75.7\%, 
respectively.
As a generative model, AsmDepictor struggles with unseen instances
when generating function symbol names, 
likely due to abnormal inlining patterns.
The underperformance of SymLM may stem from its
approach of treating function name inference as a multi-class
multi-label classification task.
Because the original work uses four architectures and incorporates
obfuscation techniques for training, our findings suggest that
achieving comparable performance requires a substantially larger
dataset than what was used in our evaluation.
}

\subsubsection{(RQ3) Inlining Impact on Malware Detection Models}
\label{ss:md}

A malware detection task aims to 
distinguish malware from benign sample(s).
We trained four traditional ML 
models: logistic regression as 
a linear model, random forest as 
an ensemble method, 
CatBoost as a gradient boosting, 
and K-nearest neighbors (KNN) 
as instance-based learning; and 
two deep learning models: 
deep neural network (DNN) and 
convolutional neural network (CNN) 
proposed by Abusnaina~\etal~\cite{dl_fhmc}.
We extract 62 statistical semantic features 
with TikNib~\cite{binkit}. 
Then, we evaluate each model 10 times using
Monte Carlo cross-validation.

\PP{Results}
\autoref{tbl:ml-detection-predication} 
presents the experimental results for 
malware detection. 
While the original models achieved 
high accuracies, 
their performance has been moderately degraded
(\eg around $20\%$) 
when tested on in-house malware samples
that are applied with extreme inlining. 
This decline illustrates how inlining introduces 
substantial changes in code structure, 
leading to misclassification.

\subsubsection{(RQ4) Inlining Impact on Malware Family Prediction Models}
\label{ss:mfp}
\begin{table*}[!htbp]
    \centering
    \caption{
    Performance comparison of  
four classical ML models (\eg 
logistic regression, random forest,
CatBoost, and k-Nearest Neighbors (KNN))
and two deep-learning-based models 
(\eg a deep neural network
and a convolutional neural network)
proposed by Abusnaina~\etal~\cite{dl_fhmc}
for malware detection (Section~\ref{ss:md})
and malware family classification (Section~\ref{ss:mfp}) tasks.
Note that we generate malware variants
with extreme inlining where
their sources have been leaked 
(Mirai~\cite{mirai-source-code} 
and Gafgyt~\cite{lizkebab-source-code}).
For both tasks, the performance of each model
significantly drops due to the static feature perturbations
(\eg call graphs, control flow graphs):
around 20\% $\downarrow$ for malware detection
and approximately 40\% $\downarrow$ for malware family prediction.
    }
    \resizebox{0.99\linewidth}{!}{
    \begin{tabular}{llrrrr|rrrr}
        \toprule
        \multirow{2}{*}{\textbf{Security Task}} & \multirow{2}{*}{\textbf{ML Model}}  & \multicolumn{4}{c}{\textbf{Malware in the Wild}} & \multicolumn{4}{c}{\textbf{Malware with Extreme Inlining}} \\
\cmidrule(lr){3-6} \cmidrule(lr){7-10}
&  & \multicolumn{1}{c}{\textbf{Accuracy}}    & \multicolumn{1}{c}{\textbf{Precision}}   & \multicolumn{1}{c}{\textbf{Recall}}   & \multicolumn{1}{c}{\textbf{F1}}  & \multicolumn{1}{c}{\textbf{Accuracy}}    & \multicolumn{1}{c}{\textbf{Precision}}   & \multicolumn{1}{c}{\textbf{Recall}}   & \multicolumn{1}{c}{\textbf{F1}}  \\ 
\midrule
\multirow{7}{*}{Malware Detection}

                                & Logistic Regression   & \cellcolor{gray!30} \textbf{0.99 ± 0.01} & \cellcolor{gray!30} \textbf{0.99 ± 0.01} & \cellcolor{gray!30} \textbf{0.99 ± 0.01} & \cellcolor{gray!30} \textbf{0.99 ± 0.01} & \cellcolor{gray!30} \textbf{0.81 ± 0.03} & \cellcolor{gray!30} \textbf{ 0.84 ± 0.02} & 0.81 ± 0.03 & \cellcolor{gray!30} \textbf{ 0.81 ± 0.03}  \\ 
                                &  Random Forest & \cellcolor{gray!30} \textbf{0.99 ± 0.01} & \cellcolor{gray!30} \textbf{0.99 ± 0.01} & \cellcolor{gray!30} \textbf{0.99 ± 0.01} & \cellcolor{gray!30} \textbf{0.99 ± 0.01} & 0.75 ± 0.13 & 0.77 ± 0.13 & 0.75 ± 0.13 & 0.74 ± 0.14  \\ 
                                 & CatBoost   & 0.98 ± 0.02 & 0.98 ± 0.02 & 0.98 ± 0.02 & 0.98 ± 0.02 & 0.77 ± 0.01 & 0.84 ± 0.00 & 0.77 ± 0.01 & 0.75 ± 0.01  \\ 
                                & KNN 	& 0.93 ± 0.01 & 0.93 ± 0.01 & 0.93 ± 0.01 & 0.93 ± 0.01 & 0.78 ± 0.00 & 0.80 ± 0.01 & 0.78 ± 0.00 & 0.77 ± 0.00  \\ 
                                 
                                  & Abusnaina \etal~\cite{dl_fhmc} (CNN) & \cellcolor{gray!30} \textbf{0.99 ± 0.01} & \cellcolor{gray!30} \textbf{0.99 ± 0.01 }& \cellcolor{gray!30} \textbf{0.99 ± 0.01} & \cellcolor{gray!30} \textbf{0.99 ± 0.01} & 0.78 ± 0.01 & 0.75 ± 0.02 & \cellcolor{gray!30} \textbf{0.83 ± 0.03} & 0.79 ± 0.01  \\ 
                                  & Abusnaina \etal~\cite{dl_fhmc} (DNN) & \cellcolor{gray!30} \textbf{0.99 ± 0.01} & \cellcolor{gray!30} \textbf{0.99 ± 0.01} & 0.99 ± 0.00 & \cellcolor{gray!30} \textbf{0.99 ± 0.01} & 0.75 ± 0.02 & 0.72 ± 0.01 & 0.82 ± 0.05 & 0.76 ± 0.02  \\

                                  \cmidrule(lr){2-10}
                                  & \textbf{Average}& 0.98 ± 0.01 & 0.98 ± 0.01 & 0.98 ± 0.01 & 0.98 ± 0.01 & 0.77 ± 0.05 & 0.78 ± 0.04 & 0.79 ± 0.01 & 0.78 ± 0.05  \\

                                  \midrule
                                  
\multirow{7}{*}{\makecell[l]{Malware Family \\ Prediction}}   
                                & Logistic Regression     & 0.88 ± 0.05 & 0.90 ± 0.06 & 0.88 ± 0.05 & 0.88 ± 0.06 & 0.48 ± 0.06 & 0.45 ± 0.07 & 0.48 ± 0.06 & 0.46 ± 0.05  \\ 
                                
                               & Random Forest & \cellcolor{gray!30} \textbf{0.91 ± 0.05} & \cellcolor{gray!30} \textbf{0.92 ± 0.05} & \cellcolor{gray!30} \textbf{0.91 ± 0.05} & \cellcolor{gray!30} \textbf{0.90 ± 0.06} & 0.50 ± 0.00 & 0.47 ± 0.06 & 0.50 ± 0.00 & 0.48 ± 0.04  \\ 
                               
                               & CatBoost & 0.89 ± 0.03 & 0.90 ± 0.03 & 0.89 ± 0.03 & 0.88 ± 0.03 & \cellcolor{gray!30} \textbf{0.53 ± 0.10} & \cellcolor{gray!30} \textbf{0.62 ± 0.30}& \cellcolor{gray!30} \textbf{0.53 ± 0.10} & \cellcolor{gray!30} \textbf{0.54 ± 0.17}  \\ 
                               & KNN    & 0.90 ± 0.06 & 0.92 ± 0.06 & 0.90 ± 0.06 & 0.90 ± 0.07 & 0.44 ± 0.05 & 0.49 ± 0.03 & 0.44 ± 0.05 & 0.46 ± 0.04  \\
                               
                                &  Abusnaina \etal~\cite{dl_fhmc} (CNN) & 0.82 ± 0.06 & 0.85 ± 0.05 & 0.82 ± 0.06 & 0.81 ± 0.06 & 0.23 ± 0.18 & 0.43 ± 0.13 & 0.23 ± 0.18 & 0.25 ± 0.14  \\ 
                                & Abusnaina \etal~\cite{dl_fhmc} (DNN) & 0.78 ± 0.07 & 0.80 ± 0.07 & 0.78 ± 0.07 & 0.76 ± 0.09 & 0.36 ± 0.15 & 0.29 ± 0.15 & 0.36 ± 0.15 & 0.31 ± 0.13  \\

                               \cmidrule(lr){2-10}
                                
                               & \textbf{Average} & 0.87 ± 0.05 & 0.88 ± 0.05 & 0.87 ± 0.05 & 0.87 ± 0.05 & 0.44 ± 0.09 & 0.46 ± 0.09 & 0.43 ± 0.09 & 0.41 ± 0.10  \\

                                        \bottomrule 

    \end{tabular}
    }
    \label{tbl:ml-detection-predication}
    \vspace{-10px}
\end{table*}

A malware family prediction 
task aims to forecast the likelihood that
an unseen malware mutation belongs 
to a known family.
Close to malware detection, we 
use the same ML-centric approaches
excluding benign applications. 
It is noted that these malware targets 
the (prevalent) 32-bit ARM architecture
in IoT devices.
To handle a class imbalance, 
we select $500$ samples per family 
with oversampling and down 
sampling~\cite{resample}.

\PP{Results}
\autoref{tbl:ml-detection-predication} demonstrates
the experimental results for malware family prediction 
(\eg 10 families). 
Using the same semantic features with 
malware detection experiments,
all ML models show
a decent performance (\eg around 87\%) 
on community-collected IoT malware. 
However, their performance significantly 
declines: \ie 40\% $\downarrow$ for 
most malware family prediction models
when tested on our in-house 
malware with extreme inlining.
Our findings indicate that all
models are sensitive 
to our extreme inlining. 
We hypothesize that this senstivity
arises from code sharing 
among IoT malware 
families~\cite{genealogy_malware}, 
where inlining obscures structural 
patterns essential for fine-grained 
classification. 
In contrast, the broader 
semantic gap in malware detection 
appears to be less affected.

\subsubsection{(RQ5) Inlining Impact on Vulnerability Detection Models}
\label{ss:vd}
A vulnerability detection task identifies
the presence of a known vulnerability
(binary classification).
Excluding vulnerability detection models 
that operate on source 
code~\cite{he2023finer,lin2017poster,
li2018vuldeepecker, le2019maximal, 
wu2022vulcnn, wang2016automatically},
we choose two binary-based
detection models~\cite{luo2023vulhawk, marcelli2022machine}
that adopt a code search (\ie probing a vulnerable function) 
by leveraging BCSD capability.
We select 12 vulnerable functions
in the libcrypto library from two firmware 
images~\cite{netgear_tplink_firmware} --
four functions from Netgear R7000 and 
eight functions from TP-Link Deco 
M4~\cite{marcelli2022machine} (\autoref{tbl:vuln_cves} in Appendix).
Accordingly, we compile two versions of the library
containing the aforementioned vulnerabilities in 
OpenSSL-1.0.2d~\cite{openssl}: one with inlining 
enabled and one without. Each version is compiled at 
six optimization levels (\cc{-O[0-3,s,z]}), resulting 
in $12,958$ and $13,298$ functions, respectively.
We rank those functions based on 
their similarity to each known vulnerable function 
and then compute the mean reciprocal rank (MRR).
We adopt MRR@100, which evaluates the Top 100 most similar functions.

\PP{Results}
\autoref{tbl_vuln_detect_mrr} presents the MRR@100 scores before and after inlining,
demonstrating the negative impact 
(\ie lower scores) on 
recognizing a vulnerability
with an average drop of 69\%.
Notably, we observe an exception with the Trex 
model~\cite{pei2020trex} at
Netgear R7000 ($0.550 \rightarrow 0.625$).
However, our further analysis reveals that 
the ranks of other similar functions decrease
except for the highest rank.

\begin{table}[t!]
    \centering
    \caption{
    Performance comparison (in MRR@100) of 
four vulnerability detection models 
with function inlining.
We select 12 vulnerable (target) functions in 
libcrypto from two firmware images~\cite{netgear_tplink_firmware} 
(Netgear R700 and TP-Link Deco-M4).
Then, we prepare a list of comparable functions 
in OpenSSL-1.0.2d~\cite{openssl}
with and without inlining for probing the vulnerable ones.
The performance of each model 
mostly significantly drops 
after inlining has been applied.
We compute the mean reciprocal rank for evaluation.
(Section~\ref{ss:vd}).
    }
    \resizebox{0.99\linewidth}{!}{
        \begin{tabular}{lrrrrr}\toprule
\multirow{2}{*}{\textbf{Model}} &\multicolumn{2}{c}{\textbf{Netgear R7000}} &\multicolumn{2}{c}{\textbf{TP-Link Deco-M4}} \\\cmidrule{2-5}
&\textbf{Non-inlining} &\textbf{Inlining} &\textbf{Non-inlining} &\textbf{Inlining} \\\midrule
\textbf{Gemini~\cite{xu2017neural}} &0.049 &0.000 &0.321 &0.004 \\
\textbf{Asm2Vec~\cite{asm2vec}} &0.256 &0.128 &0.016 &0.008 \\
\textbf{SAFE~\cite{safe}} &0.125 &0.003 &0.018 &0.003 \\
\textbf{Trex~\cite{pei2020trex}} &0.550 &0.625 &0.286 &0.048 \\
\bottomrule
\end{tabular}

    }
    \label{tbl_vuln_detect_mrr}
\end{table}

\subsection{Evaluation of Function Inlining}
\label{eval1}

\PP{Research Questions}
This section explores
the impact of function inlining
transformation itself in terms of
inlining ratios and static features,
with the following research questions.
\begin{itemize}[leftmargin=*]
    \item
    \textbf{RQ6}: How do optimization levels affect a function inlining ratio across different applications (Section~\ref{ss:fn_inl_ratio})?
    \item 
    \textbf{RQ7}: How do the 
    default compiler options (provided by clang) affect 
    a function inlining ratio (Section~\ref{ss:comp_flags_impact})? 
    \item 
    \textbf{RQ8}: Which combination of 
    compiler options (provided by opt) 
    increase a function 
    inlining ratio toward extreme inlining 
    in practice (Section~\ref{ss:lto_fn_inl})?   
    \item 
    \textbf{RQ9}: To what extent are 
    static features affected 
    by function inlining in an executable,
    such as instructions, control flow graphs, 
    and call graphs (Section~\ref{ss:static_feat})?
\end{itemize}

\begin{figure}[t!]
    \centering
    \begin{subfigure}[b]{.5\columnwidth}
        \centering
        \includegraphics[width=\linewidth]{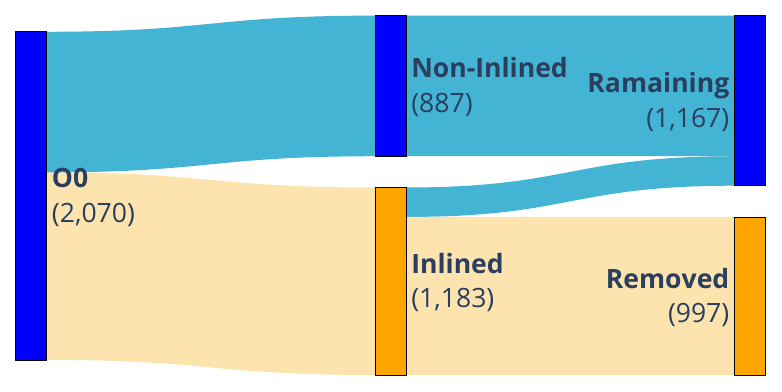}
        \caption{High optimization (\cc{-O3})}
    \end{subfigure}%
    \begin{subfigure}[b]{.5\columnwidth}
        \centering
        \includegraphics[width=\linewidth]{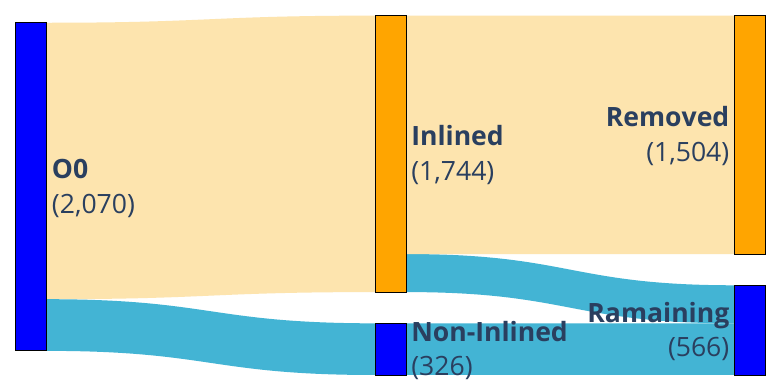}
        \caption{Extreme inlining}
    \end{subfigure}
     \vspace{-20px}
    \caption{ 
    Visualization of inlined functions
    and their presence in the coreutils package
    when compiled with (a) the \cc{-O3}
    optimization level and (b)
    our extreme inlining strategy.
    Compared to \cc{-O0}, 
    more than half of the functions have been
    inlined and yet eliminated in \cc{-O3}
    (Section~\ref{ss:fn_inl_ratio}).
    }
    \label{fig:sankey_coreutils}
     \vspace{-10px}
\end{figure}

\begin{figure*}[t!]
    \centering
    \resizebox{\linewidth}{!}{%
    \includegraphics[width=\linewidth]{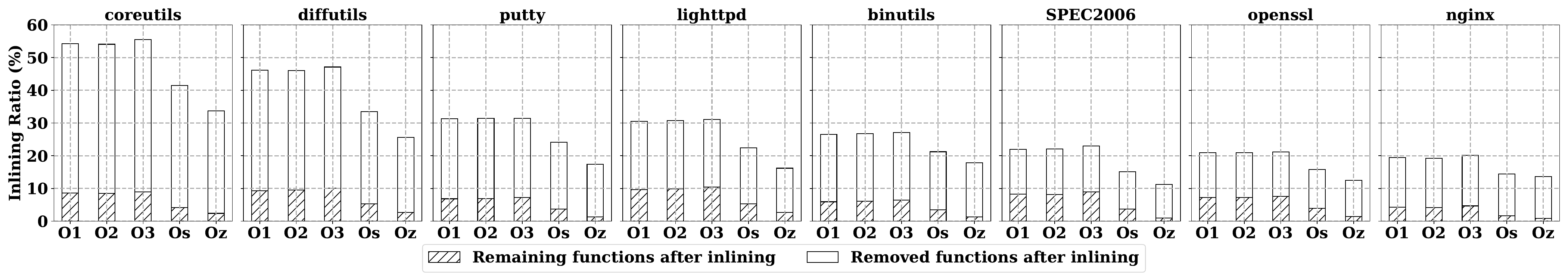}
    }
    \caption{
    Comparison of function inlining ratios compiled with different 
    optimization levels (\cc{-O[0-3,s,z]}) across eight applications.
    As the inlining decision of an individual function largely relies on
    the dynamics of a cost and a threshold by the cost model (Section~\ref{inlining_ratio}),
    inlining ratios may considerably differ.
    For example, coreutils demonstrates the highest (\eg more than half)
    in all optimization levels, whereas nginx ranks the lowest.
    The ratios in \cc{-Os} and \cc{-Oz} are relatively smaller than \cc{-O[0-3]}.
    While most inlined functions were eliminated, a non-negligible number (around 10\%) remained.
    }
    \label{fig:packages}
    \vspace{-10px} 
\end{figure*}

\subsubsection{(RQ6) Function Inlining Ratio}\label{inlining_ratio}
\label{ss:fn_inl_ratio}

\autoref{fig:sankey_coreutils} illustrates the overall flow of inlined functions 
from the coreutils binaries compiled with \cc{-O[0,3]} 
and our extreme inlining strategy
(Section~\ref{ss:lto_fn_inl}).
Starting from the whole $2,070$ functions 
(assuming little inlining with \cc{-O0}),
we observe more than half inlined functions ($1,183$).
Out of those inlined functions,
997 were eliminated, while others remained.
Driving to extreme inlining 
(more aggressive than \cc{-O3}), 
approximately two-thirds are removed.

\begin{figure}[t]
    \centering
    \resizebox{\linewidth}{!}{%
    \includegraphics[width=\linewidth]{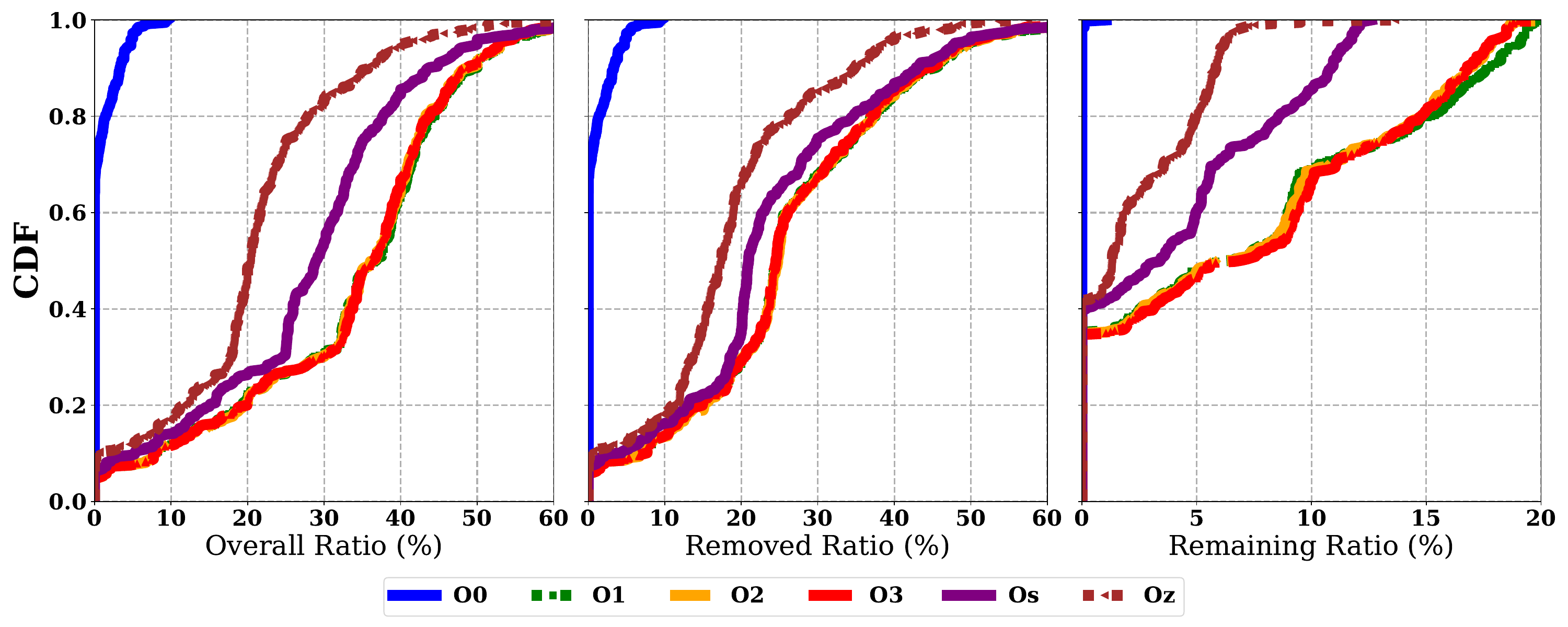}
    }
    \caption{Cumulative distribution function (CDF) of 
    function inlining ratios across all binaries in 
    our dataset %
    compiled with various optimization levels \cc{-O[0-3,s,z]} (Section~\ref{ss:fn_inl_ratio}). 
    An inlining optimization occurs even with \cc{-O0}
    while an aggressive optimization (\cc{-O3})
    drives more inlining.}
    \label{fig:binaries_cdf}
    \vspace{-10px}
\end{figure}

\PP{Function Inlining across Optimization Levels and Applications} 
\autoref{fig:packages} presents the function inlining ratio 
across eight packages in a baseline dataset with 
five optimization levels excluding \cc{-O0}. 
First, for all optimization levels, coreutils displays 
the highest inlining ratio whereas nginx ranks 
the lowest in most optimizations.
Second, in every package, \cc{-Os} and \cc{-Oz} consistently
show lower inline ratios because of the optimization purpose
to reduce a binary size (\ie the inline operation of 
a function increases a size).
Third, a function inlining ratio can vary
depending on applications.
Lastly, a non-negligible number
of inlined functions has remained (predominately) 
due to the function with an external linkage 
or even an internal linkage
that has not been inlined at all call sites.
\autoref{fig:binaries_cdf} illustrates 
the cumulative distribution 
of the inlining ratio across all binaries in our baseline dataset. 
Notably, a slight function inlining ratio 
has been observed even with \cc{-O0} 
(\eg \cc{always-inline}) shows a maximum ratio of 
9.52\% and a mean of 0.83\%
In contrast, \cc{-O[1-3]} collectively demonstrate a significantly 
high inlining ratio with a maximum of 66.67\% and 
a mean of 32.89\%, 32.77\%, and 32.70\%, respectively. 
We observe similar maximum inlining ratios 
of 65.83\% for 
\cc{-Os} and 61.67\% for \cc{-Oz}. 
However, the mean inlining ratios slightly decrease to 
27.42\% for \cc{-Os} and 20.47\% for \cc{-Oz},
indicating those optimization tactics 
stick to lowering a binary size.

\subsubsection{(RQ7) Compiler Options (provided by Clang) Impact 
on Function Inlining ratio}
\label{ss:comp_flags_impact}
\begin{figure}[t!]
    \centering
    \resizebox{1.0\linewidth}{!}{%
    \includegraphics[width=\linewidth]{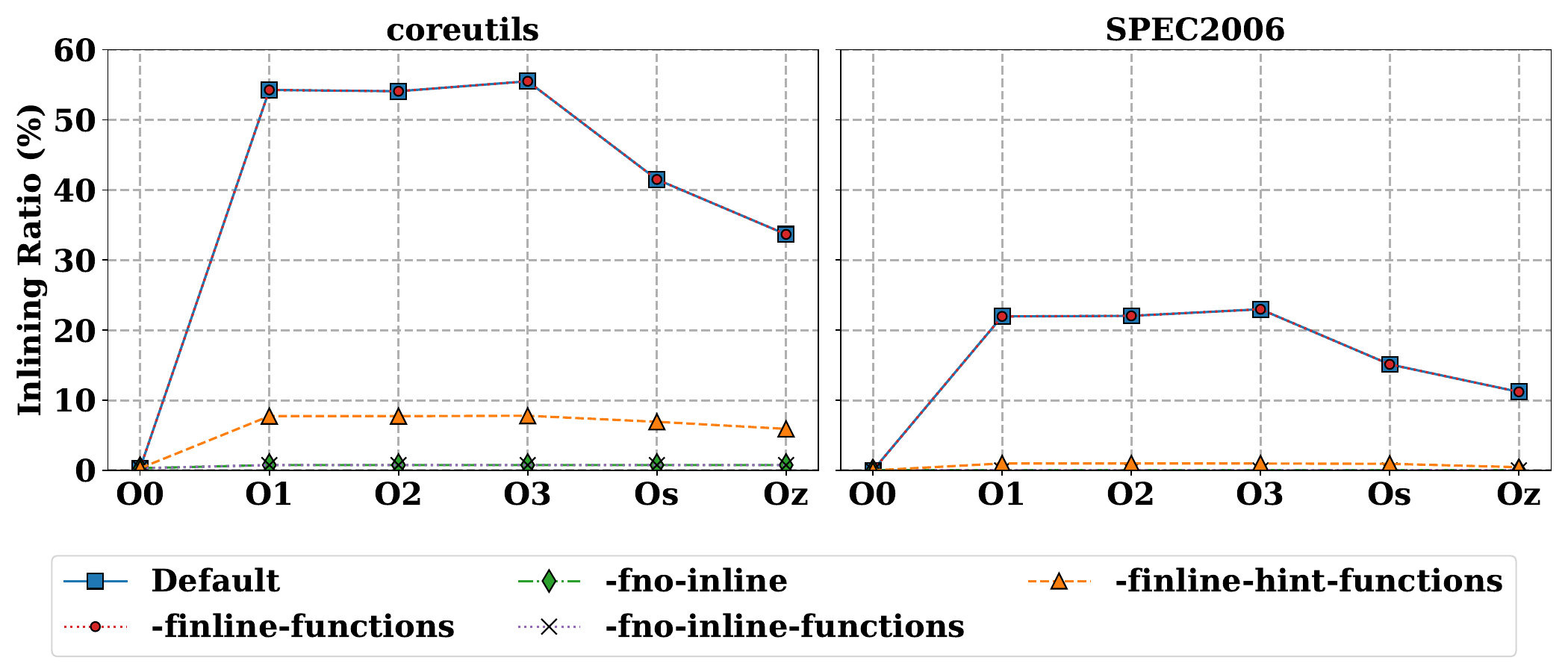}
    }
    \caption{
    Effectiveness of 
    compiler-provided options that globally affect
    the behavior of inlining (Section~\ref{ss:comp_flags_impact}).
    We measure function inlining ratios on
    coreutils and SPEC2006 
    across various optimization levels of 
    \cc{-O[0-3,s,z]}.
    We confirm the \cc{-fno-inline-functions} 
    and \cc{-fno-inline} 
    options effectively prohibit an inlining behavior.%
    }
    \label{fig:coreutils_clang_flags}
    \vspace{-10px}
\end{figure}

\autoref{fig:coreutils_clang_flags} depicts
the impact of enabling four default 
compiler options
(\cc{-finline-hint-functions}, \cc{-fno-inline-functions},
\cc{-finline-functions}, \cc{-fno-inline})
relevant to function inlining while the default
means no such options are given.
The ratio with the \cc{-finline-functions} option 
is identical to that with the default setting. 
Meanwhile, \cc{-finline-hint-functions} shows 
a relatively low inlining rate. 
In SPEC2006, the \cc{-fno-inline} and \cc{-fno-inline-functions} 
options are effective.
It is noted that \cc{always\_inline} contributes 
to maintain a small number of function inlining in coreutils
even with the inline-suppressing options.

\begin{figure}[t!]
    \centering
    \resizebox{\linewidth}{!}{%
    \includegraphics[width=\linewidth]{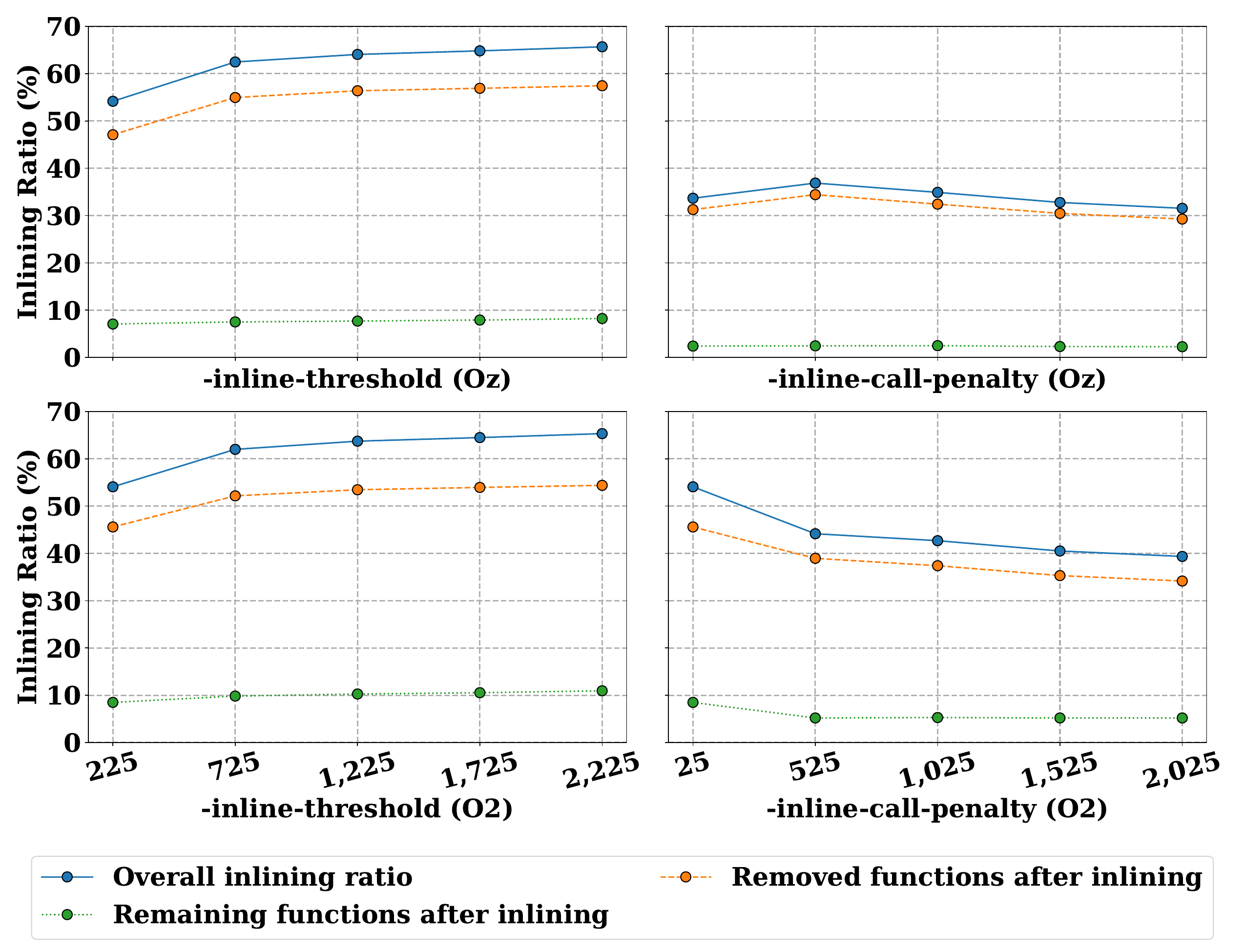}
    }
    \caption{
    Selective results of our exploration
    for seeking a combination of compiler options.
    With coreutils, we extensively tune
    (\ie increase or decrease) the values
    of the compiler options (\autoref{tbl:flags}),
    which can eventually affect function inlining.
    Our findings show that heuristically
    increasing \cc{-inline-threshold} and
    decreasing \cc{-inline-call-penalty}
    assist in growing an overall function inlining ratio (Section~\ref{ss:lto_fn_inl}).
    }    \label{fig:packages_hidden_param}
    \vspace{-10px}
\end{figure}

\subsubsection{(RQ8) Compiler Options (provided by Opt) Exploration toward Extreme Inlining}
\label{ss:lto_fn_inl}
We investigate $12$ compiler options 
associated with function inlining, aiming 
to seek a combination of the compiler options 
toward extreme inlining according to
the cost model in LLVM.
Motivated by BinTuner~\cite{binTuner}, we started
by setting a certain value that contributes to the inlining ratio.
However, we observe that adjusting the initial value could
significantly increase compilation time.
To examine the inlining ratio
within a reasonable compilation time,
we increment a specific value
by a fixed interval, such as
500 (\eg [0, 500, 1,000, ...]) 
for benign applications
and 50,000 (\eg [0, 50,000, 100,000, ...]) 
for malware. 
For an option defined as a boolean value, 
we flip the default value (\eg $true \rightarrow false$).

\PP{Results}
Compared to other options that show marginal
differences in a function inlining ratio, 
it tends to be proportional to
\cc{-inline-threshold} and inversely proportional 
to \cc{-inline-call-penalty} (\autoref{fig:packages_hidden_param}). 
The \cc{-inline-threshold} option
overrides the global threshold value while
the \cc{-inline-call-penalty} option adjusts 
the call penalty value to calculate a cost. 
For example,
the inlining ratio increases when a threshold
increases from the default value 
(\eg $225 \rightarrow 2,225$). 
Notably, these compiler options do not
bring about the heuristics
at \autoref{llvm_no_inlining}.

\begin{figure}[t!]
    \centering
    \resizebox{\linewidth}{!}{%
    \includegraphics[width=\linewidth]{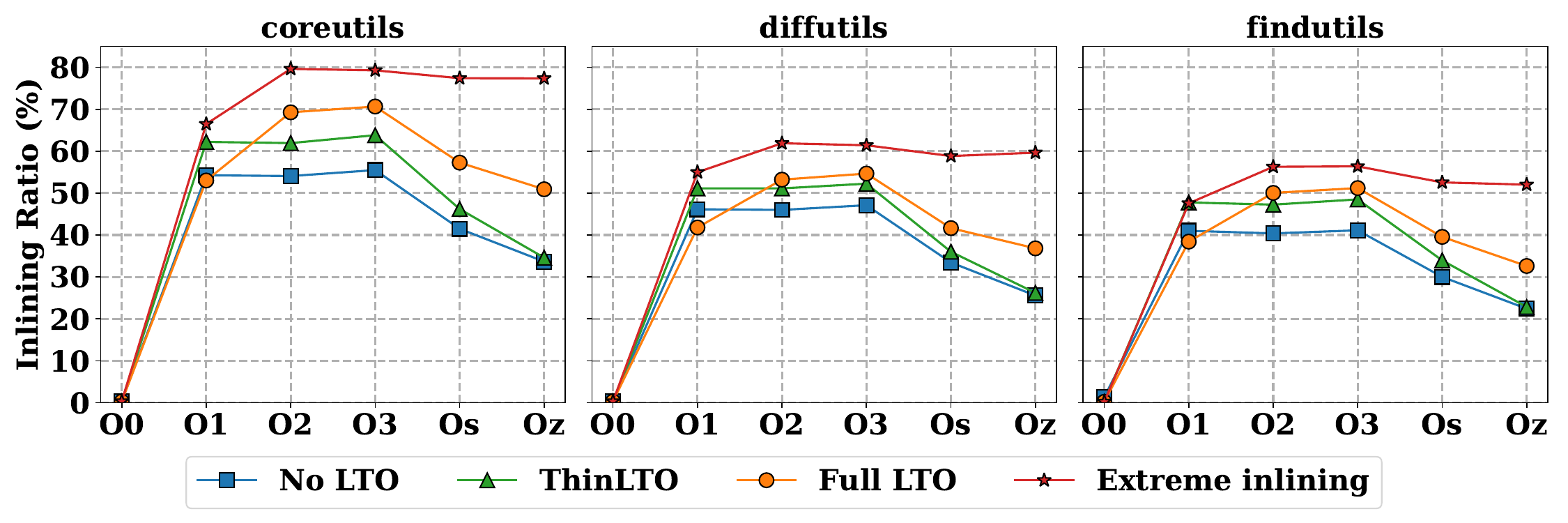}
    }
    \caption{
    Effectiveness of link time optimization (LTO) 
    across different optimization levels and applications.
    Enabling LTO clearly demonstrates a higher
    function inlining ratio at \cc{-O1} or above 
    optimization levels.
    It is worthwhile to note that our extreme 
    function inlining strategy
    even surpasses the ratio with Full LTO
    (\eg 79.64\% in coreutils).
    }
    \label{fig:packages_full_thinlto}
    \vspace{-10px}
\end{figure}

\PP{Function Inlining and LTO}
Next, we investigate how LTO affects overall 
function inlining ratios.
LTO enables an additional round of optimization 
after the initial pre-link optimization on each object file. 
We examined four scenarios: 
no LTO, Full LTO, ThinLTO, and LTO with 
our extreme function inlining. 
We set the \cc{-inline-threshold} option to 
200,000 with Full LTO,
successfully compiling three packages 
(coreutils, diffutils, and findutils). 
As in \autoref{fig:packages_full_thinlto},
Full LTO exhibits a higher inlining ratio than ThinLTO 
because Full LTO operates on a single thread 
with all object files available, 
whereas ThinLTO uses multiple threads and 
summary information. 
Interestingly, ThinLTO shows a higher inlining ratio 
than Full LTO at \cc{-O1}, potentially due to 
more efficient, aggressive optimizations. 
LTO also leverages the internalize pass, 
converting external to internal linkage 
at link time, which increases the inlining 
ratio~\cite{ThinLTO}. 
Our extreme inlining strategy achieves 
an extremely high inlining ratio, 
reaching 79.64\% in coreutils.

\subsubsection{(RQ9) Static Features Analysis According to Function Inlining}
\label{ss:static_feat}
We analyzed 62 statistical features pertaining to 
instructions, CFG, and CG using TikNib~\cite{binkit}. 
For brevity, we select 18 features with the 
highest median gap 
(\autoref{fig:instruction_cfg_cg}
in Appendix).
Note that we display the whole static feature names at
\autoref{tbl:feature_list} in Appendix.
For a fair comparison, we normalize the data 
and remove the outliers with the three-sigma 
rule~\cite{investopedia_three_sigma},
which excludes any data points that fall beyond 
three standard deviations from the mean.
We compared \cc{-O0} against extreme inlining 
across 106 binaries in coreutils. 
\autoref{fig:instruction_cfg_cg} 
in Appendix shows that 
extreme inlining drastically changes 
the statistical features of an executable binary. 
For example, there is a significant increase 
in arithmetic instructions (Features 6 and 8), likely 
due to the inlining of functions with arithmetic operations 
across multiple call sites. 
Additionally, we observe a decrease in the 
number of loops (Features 38 and 39) but an increase 
in loop size (Features 46, 48, and 49). 
This suggests that inlining results in fewer but larger loops. 
Extreme inlining may enable more aggressive optimizations, 
such as loop unrolling and removing smaller loops. 
These changes significantly impact 
the statistical features, which we expect 
will affect ML models relying on such features.

\section{Implementation}
\label{sec:implem}

\PP{Ground Truth Extraction for Executable Binaries} 
We compile varying software in the ELF (Executable and 
Linkable Format) format with the \cc{-g} option, which
generates debugging information.
We develop a script in Python
with the \cc{pyelftools} library~\cite{pyelftools}
to parse ELF and extract the \cc{DW\_AT\_inline} 
attribute (\ie inlined function) in a function symbol from
DWARF~\cite{dwarf5}.
Identifying a list of callers with the
\cc{DW\_TAG\_subprogram} tag at each function,
we traverse child nodes with the \cc{DW\_TAG\_inlined\_subroutine} 
tag in each DWARF entry, which indicates that 
a function contains multiple inlined functions. 
Besides, we leverage TikNib~\cite{tiknib} to extract
$62$ features (\autoref{tbl:feature_list} in Appendix) 
that can impact a control flow graph,
a call graph, and instruction sequences.

\PP{ML-based Models}
First, we leverage the BCSD 
benchmark implementations
~\cite{marcelli2022machine}
to assess a BCSD task
instead of reinventing the wheel,
including Asm2Vec~\cite{asm2vec},  
Gemini~\cite{xu2017neural}, Trex~\cite{pei2020trex}, 
and SAFE~\cite{safe}. 
We adopt the original implementation
of BinShot~\cite{binshot}, which is
unavailable on the above benchmark.
Second, for malware detection and family
prediction, we follow
the Lei~\etal~\cite{ml_detection_family_hyper}'s
approach to generate 
four traditional ML models 
(\autoref{tbl:ml-detection-predication})
with the Scikit-learn library~\cite{ScikitLearn}. 
Additionally, we re-implement 
two deep learning models proposed by Abusnaina \etal \cite{Abusnaina} with TensorFlow~\cite{tensorflow2} (\autoref{tbl:ml-detection-predication}). 
Third, we retrained AsmDepictor~\cite{asmdepictor} 
and SymLM~\cite{jin2022symlm}
with our dataset compiled in LLVM
for function name prediction.
Fourth, we assess vulnerability detection 
by leveraging the BCSD benchmark 
implementation~\cite{marcelli2022machine} and available vulnerable firmware corpus~\cite{netgear_tplink_firmware}.

\section{Discussion and Future Work}

\PP{Unreported Function Inlining Cases in DWARF}
Although DWARF holds plentiful information about 
function inlining, there are several cases that 
(deliberately) do not report them.
We observe missing the record of inlining 
when the whole inlined function body has been removed (\eg dead code elimination) during transformations.
Another unreported case is when a specific intrinsic 
function
has been inlined into other functions.
This may cause a discrepancy between 
the output of the function inliner pass (\eg \cc{-Rpass=inline})
and that of parsing DWARF.

\PP{Dataset Representativeness} 
Although we carefully select a wide range
of different applications as a dataset
for our experiments, it may not sufficiently
represent all binaries to safely generalize
our findings.
We leave more scenarios and
edge cases as our future work, which
need to be explored with
a diverse dataset to unearth
the behavior of inlining a function.

\begin{figure}[t]
    \centering
    \resizebox{\linewidth}{!}{%
    \includegraphics[width=\linewidth]{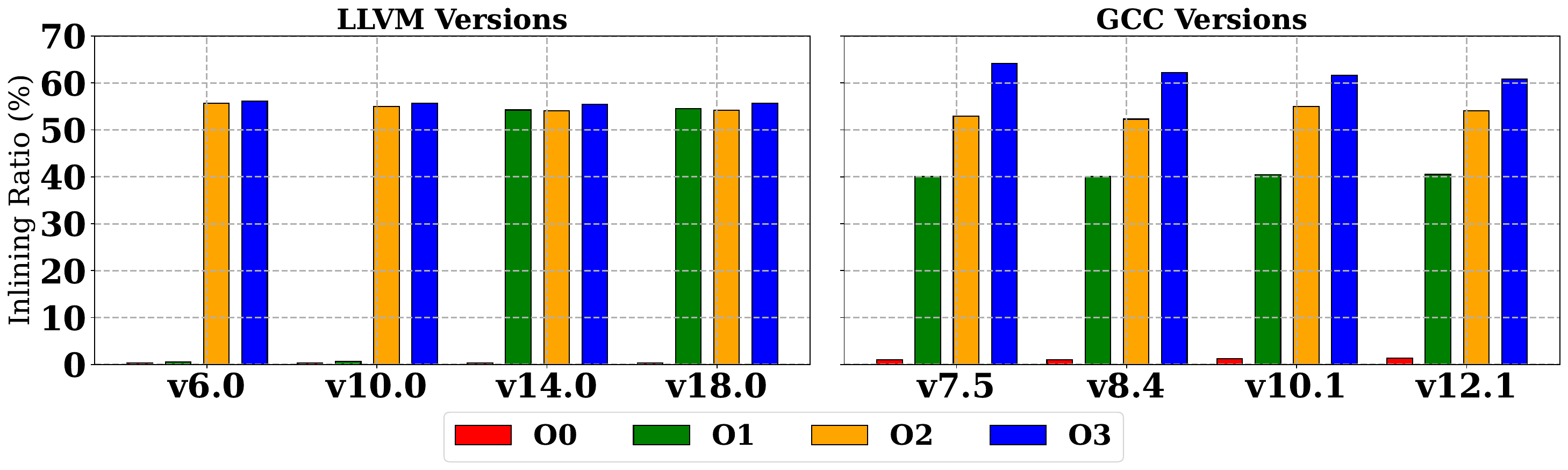}
    }
    \caption{
    Comparison of function inlining ratios across
    different optimization levels (O0–O3) for 
    multiple versions of LLVM (left) and 
    GCC (right). 
    Higher optimization levels 
    consistently yield higher inlining ratios across 
    all compiler versions, with GCC exhibiting 
    more variability between versions 
    compared to LLVM.
    }
    \label{fig:across_compilers}
    \vspace{-15px}
\end{figure}

\PP{Behavioral Differences between Compilers} 
Empirically, it is possible to 
have inconsistent outcomes
due to the dynamics of function-inlining for 
code optimization across compilers and compiler versions. 
\autoref{fig:across_compilers} presents
the function inlining ratios 
across different versions of 
GCC and LLVM compiler at 
various optimization levels.
While LLVM exhibits relatively 
consistent inlining behaviors,
GCC shows more aggressive inlining,
particularly at higher optimization levels.
Although our experiments focus on a specific version of LLVM,
our observation remains applicable across its
versions, providing valuable insights into how modern compilers
handle function inlining. 
A deeper investigation into GCC’s internal 
inlining decisions is left for future work.

\PP{Evaluation Metric for Function Inlining Ratio} 
We utilize the evaluation metric with 
the ratio of function inlining 
based on the number of (inlined) functions. 
However, an inlining decision is
made at the call site granularity that takes
a caller and a callee into account.
Hence, the case of a single function that 
has been inlined into multiple locations or
a chain of (nested) inlining may not been
represented with our metric.

\PP{\UP{Potential Mitigations}}
\UP{
We propose several potential mitigations: \WC{1} augmenting 
model training with inlined data, \WC{2} 
employing adversarial training that incorporates 
compiler-aware transformations, and 
\WC{3} applying inlining-aware pre-processing 
(e.g., Highliner~\cite{Highliner} uses de-inlining 
heuristics).
We direct interested readers to Appendix~\autoref{tbl:ml-detection-augmention}
for preliminary results on augmented training 
with inlined functions.
}

\section{Related Work}

\PP{Feature Engineering for Binary Reversing}
Genius~\cite{feng2016scalable} utilizes an attributed CFG (ACFG)
where they incorporate eight features (statistical and structural) 
at the basic block granularity.
Similarly, DL-FHMC~\cite{dl_fhmc} introduces 23 additional features on top of CFG.
Additionally, ImOpt~\cite{jiang2020similarity} explores features at the IR level to tackle
compiler optimization and obfuscation techniques. 
In contrast, $\alpha$Diff~\cite{liu2018alphadiff} 
utilizes a deep neural network
to directly extract features.
TikNib~\cite{binkit} extracts 
72 statistical features 
using a large-scale benchmark, which
demonstrates that simple interpretable models
show comparable performance to state-of-the-art 
deep learning models with precise features.
Note that we borrow 62 applicable features 
at the binary level to investigate the 
impact of inlining.

\PP{Function Inlining and Binary} 
Several prior works deal with 
function inlining from engineering
aspects.
Damásio~\etal~\cite{inling_code_size} focus on
inlining for code size reduction, while 
Theodoridis~\etal~\cite{exploiting_inlining}
explore optimal inlining strategies.
In parallel, other studies examine function 
inlining in the context of binary analysis.
One close effort to our work is
Jia~\etal~\cite{1to1_1ton}, who investigate
the impact of function inlining 
on binary similarity analysis.
Bingo~\cite{bingo} introduces 
a dynamic inlining-simulation 
strategy that recursively 
expands callee functions 
to improve similarity detection.
Similarly, Asm2Vec~\cite{asm2vec} adapts the Bingo's strategy 
for static analysis using selective callee expansion. 
FSmell~\cite{fsmell} proposes a ML–based 
framework to detect inlined functions through
instruction topology graphs.
Meanwhile, Koo \etal~\cite{lookback_function} and
AsmDepictor~\cite{asmdepictor} emphasize 
the importance of a deeper understanding 
of function inlining.
Nonetheless, many ML-based studies~\cite{cross_arch_bug_search, 
                    discovRE, 
                    Inlining-vulunerbility, 
                    extracting_cross_platform,
                    xue2019machine, 
                    jin2022symlm, 
                    patrick2020probabilistic, 
                    katz2018using, XDA, BYTEWEIGHT}
for binary analysis
tend to underestimate or overlook
inlining effects without thorough investigation.
Unlike previous approaches, 
our work demystifies the compiler's 
inline decision process 
and the possibility of misuse by deliberately crafting evasive 
binary mutations through extreme inlining.

\PP{ML-assisted Approaches for Static Binary Analysis} 
Over the past decade, the widespread adoption of ML-assisted
security tasks across various fields have demonstrated promising results. 
Such examples include 
malware detection, malware family classification, BCSD, and function name prediction tasks. 
In the area of BCSD, VulSeeker~\cite{vulseeker} employs a 
Siamese network-based graph embedding model to enhance similarity detection.
InnerEye \cite{innereye} and SAFE \cite{safe} adopt 
natural language processing (NLP) approaches for learning 
semantics from assembly code.
In a similar vein, Asm2Vec~\cite{asm2vec} and DeepBinDiff~\cite{deepbindiff} 
utilize unsupervised learning for training in the context of instructions, thereby 
improving their ability to detect code similarities.
Lately, BinShot~\cite{binshot} learns a weighted distance vector with 
to better take dissimilar codes apart, which we include for our evaluation.
For malware applications, Alasmary \etal~\cite{android_mal}
conduct a comprehensive CFG-based IoT malware detection task 
on Android.
Cozzi \etal~\cite{genealogy_malware} examine the lineage of malware families, 
tracking their relationships across variants.
Furthermore, BinTuner~\cite{binTuner} and CARDINAL~\cite{malware_complier} attempt
to re-compile malware samples
with different compilation option settings
to defeat signature-based approaches 
that rely on static engineering features. 
Meanwhile, recent advances in an attempt to recover lost 
information during compilation presents the inference
of function and variable names, types, and even decompiled code.
Debin~\cite{debin} focuses on recovering 
variable names and types that learn the underlying semantics
from instruction sequences at the binary level.
NERO~\cite{nero} trains enriching representations of call sites 
based on augmented control flow graphs to predict function symbol names.
Similarly, AsmDepictor~\cite{asmdepictor} leverages
the state-of-the-art Transformer-based model into a function name prediction task,
which is included in our evaluation.

\PP{\UP{Study on Compiler Flags and Behaviors}}
\UP{Dong \etal~\cite{dong2015studying} and Zhang~\cite{zhang2024compiler} \etal 
study symbolic execution to explore compiler flags: 
the former compare 
different optimization levels, while
the latter empirically analyze 
GCC/Clang optimization flags. 
Ren \etal~\cite{ren2025revisiting} focus on LLVM peephole 
optimizations 
for binary diffing, and 
BinTuner~\cite{binTuner} explores large compiler 
flag spaces for binary similarity. 
Meanwhile, our work targets ML-based binary analysis 
and moves toward extreme inlining to evaluate and 
evade ML models by systematically examining 
LLVM’s inlining code base.}

\section{Conclusion}

Function inlining is a well-known optimization technique by modern compilers.
Such an inlining behavior can considerably affect 
static features for a binary reversing task; 
however, it has
yet well explored despite its importance and impact.
In this work, we first conduct a comprehensive study
focusing on function inlining, 
including (but not limited to)
the investigation of its decision pipeline, 
compiler options that directly affect 
an inlining ratio, unearthing common misbeliefs, and
the evaluation of security implications
on ML-oriented applications.
Our major findings indicate that 
function inlining 
can be exploited for malicious purposes, 
which requires paying attention
when building ML-based models.

\section*{Acknowledgments}
We thank the anonymous reviewers 
for their constructive feedback.
This work was partially 
supported by the grants from
Institute of Information \& Communications 
Technology Planning \& Evaluation (IITP),
funded by the Korean government 
(MSIT; Ministry of Science and ICT): 
No. RS-2024-00337414,  
No. RS-2024-00437306,
No. RS-2020-II201821, and
No. RS-2022-II220688.
Additional support was provided by  the Basic Science Research Program through
    the National Research Foundation of Korea (NRF),
funded by the Ministry of Education
	of the Government of South Korea 
	: No. RS-2025-02293072.
	Any opinions, findings, and conclusions or 
	recommendations expressed in
	this material are those of the authors and 
	do not necessarily reflect
	the views of the sponsor.

\newpage

\bibliographystyle{IEEEtranS} 
\bibliography{references}

\appendices
\section{Supplementary Experiments and Analyses} 
\label{s:appendix}

\subsection{DWARF Information for Function Inlining}

\begin{table}[H]
    \centering
    \caption{Debugging information
    fields in DWARF~\cite{dwarf5}
    associated with function inlining.
    }
    \resizebox{0.99\columnwidth}{!}{
    \begin{tabular}{lll}
        \toprule
        \textbf{Name} & \textbf{Type} & \textbf{Description} \\ \midrule
         DW\_TAG\_inlined\_subroutine & T & Particular inlined instance of a (sub)routine \\
         DW\_AT\_abstract\_origin & A & Pointer to the DIE of the inlined subprogram \\
         DW\_AT\_inline & A & Constant describing inlining behavior \\
         DW\_AT\_decl\_file & A & File containing source declaration \\
         DW\_AT\_low\_pc & A & Low address of a machine code \\
         DW\_AT\_high\_pc & A & High address of a machine code \\
         DW\_AT\_call\_file & A & File containing an inlined subroutine call \\
         DW\_AT\_call\_line & A & Line number of an inlined subroutine call \\
         DW\_AT\_call\_column & A & Column position of an inlined subroutine call  \\ \bottomrule

    \end{tabular}
    }
    \label{tab:DIEtagat}
\end{table}

\subsection{Code Semantics on Vulnerability Detection}
\label{vuln_appendix_inl}

\autoref{tbl:vuln_cves} displays CVEs and corresponding vulnerable functions described in Section~\ref{ss:vd}.
For example, with Trex~\cite{pei2020trex}, we discover that 
the similarity rank of \cc{BN\_dec2bn} has increased
after being inlined into \cc{BN\_asc2bn}:~\eg $1,044 \rightarrow 14$,
which indicates that the model recognizes the code semantics.
As such, identifying 
the semantics of a vulnerability within 
the inlining function 
(\ie containing a vulnerable function)
is a good sign for detection.
However, based on our observation, 
the mixture of different 
code semantics could
degrade the model's performance
by additional optimizations 
and nested-inlining may significantly.

\begin{table}[H]
    \centering
    \caption{
    The specific names and CVEs of the vulnerable functions we explored in Section~\ref{ss:vd}.
    }
    \resizebox{0.8\linewidth}{!}{
        \begin{tabular}{lllr}\toprule
&\textbf{CVE} &\textbf{Query Function} \\\midrule
\textbf{} &\textbf{CVE-2016-2182} &\cc{BN\_bn2dec} \\
\textbf{Netgear} &\textbf{CVE-2019-1563} &\cc{CMS\_decrypt} \\
\textbf{R7000} &\textbf{CVE-2016-6303} &\cc{MDC2\_Update} \\
\textbf{} &\textbf{CVE-2019-1563} &\cc{PKCS7\_dataDecode} \\\midrule
\textbf{} &\textbf{CVE-2016-2182} &\cc{BN\_bn2dec} \\
\textbf{} &\textbf{CVE-2016-0797} &\cc{BN\_dec2bn} \\
\textbf{} &\textbf{CVE-2016-0797} &\cc{BN\_hex2bn} \\
\textbf{TP-Link} &\textbf{CVE-2019-1563} &\cc{CMS\_decrypt} \\
\textbf{Deco-M4} &\textbf{CVE-2016-2105} &\cc{EVP\_EncodeUpdate} \\
\textbf{} &\textbf{CVE-2019-1563} &\cc{PKCS7\_dataDecode} \\
\textbf{} &\textbf{CVE-2016-0798} &\cc{SRP\_VBASE\_get\_by\_user} \\
\textbf{} &\textbf{CVE-2016-2176} &\cc{X509\_NAME\_oneline} \\
\bottomrule
\end{tabular}

    }
    \label{tbl:vuln_cves}
\end{table}

\begin{table}[t!]
    \centering
    \caption{Descriptions of static
    features in an executable binary from Binkit~\cite{binkit}.
    The 36 features (1-36) depict 
    instruction-level features, while
    the 20 (37-56) and six (57-62) features
    are relevant to 
    a control flow graph and a call graph, respectively.
    We analyze how each static feature has been affected by 
    extreme inlining.
    }
    \resizebox{1\linewidth}{!}{
    \begin{tabular}{ll}
        \toprule
        \textbf{Index} & \textbf{Feature Description}   \\ 
\midrule
1                                                         & Number of instructions                                 \\ 
2                                                         & Average number of instructions                               \\ 
3                                                         & Number of unknown instructions                               \\ 
4                                                         & Average number of unknown instructions                       \\ 
5                                                         & Number of absolute arithmetic instructions                   \\ 
6                                                         & Average number of absolute arithmetic instructions           \\ 
7                                                         & Number of arithmetic instructions                            \\ 
8                                                         & Average number of arithmetic instructions                    \\ 
9                                                         & Number of comparison instructions                            \\ 
10                                                        & Average number of comparison instructions                    \\ 
11                                                        & Number of absolute control transfer instructions             \\ 
12                                                        & Average number of absolute control transfer instructions     \\ 
13                                                        & Number of conditional control transfer instructions          \\ 
14                                                        & Average number of conditional control transfer instructions  \\ 
15                                                        & Number of group jump instructions                            \\ 
16                                                        & Average number of group jump instructions                    \\ 
17                                                        & Number of absolute data transfer instructions                \\ 
18                                                        & Average number of absolute data transfer instructions        \\ 
19                                                        & Number of data transfer instructions                         \\ 
20                                                        & Average number of data transfer instructions                 \\ 
21                                                        & Number of control transfer instructions                      \\ 
22                                                        & Average number of control transfer instructions              \\ 
23                                                        & Number of group call instructions                            \\ 
24                                                        & Average number of group call instructions                    \\ 
25                                                        & Number of group return instructions                          \\ 
26                                                        & Average number of group return instructions                  \\ 
27                                                        & Number of miscellaneous instructions                         \\ 
28                                                        & Average number of miscellaneous instructions                 \\ 
29                                                        & Number of shift instructions                                 \\ 
30                                                        & Average number of shift instructions                         \\ 
31                                                        & Number of logic instructions                                 \\ 
32                         & Average number of logic instructions                                             \\
33                         & Number of bit flag instructions                                                  \\
34                         & Average number of bit flag instructions                                          \\
35                         & Number of floating-point instructions                                            \\
36                         & Average number of floating-point instructions                                    \\ \midrule
37                         & Size of CFG                                                   \\
38                         & Number of loops in a CFG                                        \\
39                         & Number of interprocedural loops in a CFG                        \\
40                         & Number of strongly connected components in a CFG                \\
41                         & Number of back edges in a CFG                                   \\
42                         & Number of breadth-first search edges in a CFG                   \\
43                         & Maximum width of CFG                                          \\
44                         & Maximum depth of CFG                                          \\
45                         & Sum of the sizes of all loops in a CFG                          \\
46                         & Sum of the sizes of all interprocedural loops in a CFG          \\
47                         & Sum of the sizes of all strongly connected components in a CFG  \\
48                         & Average size of loops in a CFG                                  \\
49                         & Average size of interprocedural loops in a CFG                  \\
50                         & Average size of strongly connected components in a CFG          \\
51                         & Number of incoming edges in a CFG                               \\
52                         & Number of outgoing edges in a CFG                               \\
53                         & Total number of edges (in-degree + out-degree) in a CFG         \\
54                         & Average number of incoming edges in a CFG                       \\
55                         & Average number of outgoing edges in a CFG                       \\
56                         & Average number of edges (in-degree + out-degree) in a CFG       \\ \midrule
57                         & Number of caller functions in a CG                                     \\
58                         & Number of callee functions in a CG                                     \\
59                         & Number of imported callee functions in a CG                            \\
60                         & Number of incoming calls in a CG                                       \\
61                         & Number of outgoing calls in a CG                                       \\
62                         & Number of imported calls in a CG                                       \\ 

\bottomrule
    \end{tabular}
    }
    
    \label{tbl:feature_list}
\end{table}

\begin{table*}[t!]
\centering
\caption{
\UP{
Summary of extreme inlining recipes 
for each task (T1–T5). 
}
}
\resizebox{0.99\linewidth}{!}{
    \begin{tabular}{ll|l|l|l|l}
        \toprule
        
\multirow{2}{*}{\textbf{Type}} &
\multirow{2}{*}{\textbf{Package or}} &
\multicolumn{4}{c}{\textbf{Extreme Inlining Recipe Settings}} \\
\cmidrule(lr){3-6}
 & \textbf{Application(s)} & 
\textbf{T1–2, T5} & 
\textbf{T3} & 
\textbf{T4} & 
\textbf{T3* (augmentation variants)} \\
\midrule

\multirow{13}{*}{\textbf{Benign}} & coreutils & -Oz -inline-threshold=2225 &
  -O3 -flto=full; -O3 -flto=full -inline-threshold=200000 &
  - &
  -O3 -inline-threshold=2225 \\

 & binutils & -O3 & - & - & - \\

 & diffutils & -O3 -flto=full &
  -Oz -inline-threshold=2225 &
  - &
  -O3 -inline-threshold=2225 \\

 & findutils & -O3 -flto=full &
  -Oz -inline-threshold=2225 &
  - &
  -O3 -inline-threshold=2225 \\

 & openssl & -O3 -flto=full &
  -Oz -inline-threshold=2225 &
  - &
  -O3 -inline-threshold=2225 \\

 & lvm2 & -O3 -flto=full & - & - & - \\

 & gsl & -Oz -inline-threshold=2225 &
  -O3 -flto=full &
  - &
  -O3 -inline-threshold=2225 \\

 & valgrind & -Oz -inline-threshold=2225 &
  - &
  - &
  -O3 -inline-threshold=2225 \\

 & openmpi & -O3 -flto=full &
  -Oz -inline-threshold=2225 &
  - &
  -O3 -inline-threshold=2225 \\

 & putty & -O3 -flto=full &
  -Oz -inline-threshold=2225 &
  - &
  -O3 -inline-threshold=2225 \\

 & nginx & -Oz -inline-threshold=2225 &
  -O3 -flto=full &
  - &
  -O3 -inline-threshold=2225 \\

 & lighttpd & -O3 -flto=full &
  -Oz -inline-threshold=2225 &
  - &
  -O3 -inline-threshold=2225 \\

\midrule

\multirow{2}{*}{\textbf{Malware}} & Mirai & - &
  \multicolumn{2}{l|}{-O(1\textbar{}2\textbar{}3\textbar{}s\textbar{}z) --inline-threshold=(225\textbar{}500225)} &
  -O(1\textbar{}2\textbar{}3\textbar{}s\textbar{}z) --inline-threshold=(225\textbar{}2000) \\

 & Gafgyt & - &
  \multicolumn{2}{l|}{-O(1\textbar{}2\textbar{}3\textbar{}s\textbar{}z) --inline-threshold=(225\textbar{}500225) -flto=full} &
  -O(1\textbar{}2\textbar{}3\textbar{}s\textbar{}z) --inline-threshold=(225\textbar{}2000) -flto=full \\

\bottomrule

    \end{tabular}
}
\label{tab_extreme_inlining_recipe}
\vspace{-20px}
\end{table*}

\begin{table*}[!t]
    \centering
    \caption{
    \UP{Performance comparison on the malware 
    detection task (T3) with newly generated 
    extreme-inlining variants.
    Interestingly, augmenting the 
    training set improves 
    neural models (\eg CNN, DNN), 
    while non-neural models (\eg
    logistic regression, random forest, CatBoost) show little benefit.}
    }
    \resizebox{0.99\linewidth}{!}{
    \begin{tabular}{llrrrr|rrrr}
        \toprule
        \multirow{2}{*}{\textbf{Training Setup}} & \multirow{2}{*}{\textbf{ML Model}}  
& \multicolumn{4}{c}{\textbf{Malware in the Wild}} 
& \multicolumn{4}{c}{\textbf{Malware with Extreme Inlining}} \\
\cmidrule(lr){3-6} \cmidrule(lr){7-10}
& & \textbf{Accuracy} & \textbf{Precision} & \textbf{Recall} & \textbf{F1} 
  & \textbf{Accuracy} & \textbf{Precision} & \textbf{Recall} & \textbf{F1} \\
\midrule

\multirow{9}{*}{\textbf{Before Augmentation}}
& Logistic Regression 
    & \cellcolor{gray!30}\textbf{0.99 ± 0.00} 
    & \cellcolor{gray!30}\textbf{0.99 ± 0.00} 
    & 0.99 ± 0.01 
    & \cellcolor{gray!30}\textbf{0.99 ± 0.00}
    & 0.60 ± 0.02 
    & 0.64 ± 0.03 
    & 0.47 ± 0.05 
    & 0.54 ± 0.03 \\

& Random Forest       
    & \cellcolor{gray!30}\textbf{0.99 ± 0.00} 
    & \cellcolor{gray!30}\textbf{0.99 ± 0.00} 
    & \cellcolor{gray!30}\textbf{0.99 ± 0.00} 
    & \cellcolor{gray!30}\textbf{0.99 ± 0.00}
    & 0.78 ± 0.01 
    & 0.70 ± 0.02 
    & 0.95 ± 0.03 
    & 0.81 ± 0.02 \\

& CatBoost            
    & 0.98 ± 0.00 
    & \cellcolor{gray!30}\textbf{0.99 ± 0.00} 
    & 0.97 ± 0.01 
    & 0.98 ± 0.00 
    & \cellcolor{gray!30}\textbf{0.78 ± 0.01} 
    & 0.70 ± 0.01 
    & \cellcolor{gray!30}\textbf{0.97 ± 0.01} 
    & \cellcolor{gray!30}\textbf{0.82 ± 0.01} \\

& KNN                 
    & \cellcolor{gray!30}\textbf{0.99 ± 0.00} 
    & \cellcolor{gray!30}\textbf{0.99 ± 0.00} 
    & 0.99 ± 0.01 
    & \cellcolor{gray!30}\textbf{0.99 ± 0.00}
    & 0.68 ± 0.04 
    & 0.69 ± 0.03 
    & 0.65 ± 0.09 
    & 0.67 ± 0.06 \\

\cmidrule(lr){2-10}

& \textbf{Average (Non-Neural)}    
    & 0.99 ± 0.00
    & 0.99 ± 0.00
    & 0.99 ± 0.00
    & 0.99 ± 0.00
    & 0.71 ± 0.08
    & 0.68 ± 0.03
    & 0.76 ± 0.20
    & 0.71 ± 0.11 \\
    
\cmidrule(lr){2-10} 

& CNN                 
    & 0.99 ± 0.01 
    & 0.99 ± 0.01 
    & 0.99 ± 0.01 
    & 0.99 ± 0.01
    & 0.75 ± 0.05 
    & \cellcolor{gray!30}\textbf{0.76 ± 0.09} 
    & 0.77 ± 0.10 
    & 0.76 ± 0.04 \\

& DNN                 
    & \cellcolor{gray!30}\textbf{0.99 ± 0.00}
    & \cellcolor{gray!30}\textbf{0.99 ± 0.00}
    & \cellcolor{gray!30}\textbf{0.99 ± 0.00}
    & \cellcolor{gray!30}\textbf{0.99 ± 0.00}
    & 0.71 ± 0.04 
    & 0.73 ± 0.03
    & 0.67 ± 0.10 
    & 0.70 ± 0.06 \\

\cmidrule(lr){2-10} 
& \textbf{Average (Neural)}    
    & 0.99 ± 0.01
    & 0.99 ± 0.01
    & 0.99 ± 0.01
    & 0.99 ± 0.01
    & 0.73 ± 0.03
    & 0.75 ± 0.02
    & 0.72 ± 0.05
    & 0.73 ± 0.03 \\

\midrule

\multirow{9}{*}{\textbf{After Augmentation}}
& Logistic Regression 
    & 0.96 ± 0.01 
    & \cellcolor{gray!30}\textbf{0.99 ± 0.01} 
    & 0.94 ± 0.02 
    & 0.96 ± 0.01
    & 0.64 ± 0.07 
    & 0.98 ± 0.02 
    & 0.28 ± 0.14 
    & 0.42 ± 0.17 
    
    \\

& Random Forest       
    & 0.98 ± 0.01 
    & \cellcolor{gray!30}\textbf{0.99 ± 0.01} 
    & 0.97 ± 0.01 
    & 0.98 ± 0.01
    & 0.69 ± 0.02 
    & 0.90 ± 0.00 
    & 0.38 ± 0.03 
    & 0.50 ± 0.02 \\

& CatBoost            
    & 0.98 ± 0.01 
    & \cellcolor{gray!30}\textbf{0.99 ± 0.01} 
    & 0.96 ± 0.02 
    & 0.98 ± 0.01
    & \cellcolor{gray!30}\textbf{0.84 ± 0.02} 
    & 0.99 ± 0.01 
    & \cellcolor{gray!30}\textbf{0.69 ± 0.03} 
    & \cellcolor{gray!30}\textbf{0.81 ± 0.02} \\

& KNN                 
    & \cellcolor{gray!30}\textbf{0.99 ± 0.00} 
    & \cellcolor{gray!30}\textbf{0.99 ± 0.01} 
    & \cellcolor{gray!30}\textbf{0.99 ± 0.01} 
    & \cellcolor{gray!30}\textbf{0.99 ± 0.00}
    & 0.81 ± 0.01 
    & \cellcolor{gray!30}\textbf{0.99 ± 0.00} 
    & 0.62 ± 0.03 
    & 0.77 ± 0.02 \\

\cmidrule(lr){2-10}

& \textbf{Average (Non-Neural)}    
    & 0.98 ± 0.01
    & 0.99 ± 0.00
    & 0.97 ± 0.02
    & 0.98 ± 0.01
    & 0.75 ± 0.09
    & 0.97 ± 0.05
    & 0.49 ± 0.18
    & 0.63 ± 0.18 \\

\cmidrule(lr){2-10}

& CNN                 
    & \cellcolor{gray!30}\textbf{0.99 ± 0.00} 
    & \cellcolor{gray!30}\textbf{0.99 ± 0.01} 
    & 0.98 ± 0.01 
    & \cellcolor{gray!30}\textbf{0.99 ± 0.00}
    & 0.84 ± 0.02 
    & \cellcolor{gray!30}\textbf{0.99 ± 0.00} 
    & 0.68 ± 0.05 
    & 0.81 ± 0.03 \\

& DNN                 
    & \cellcolor{gray!30}\textbf{0.99 ± 0.00} 
    & \cellcolor{gray!30}\textbf{0.99 ± 0.01} 
    & 0.98 ± 0.01 
    & \cellcolor{gray!30}\textbf{0.99 ± 0.00}
    & 0.81 ± 0.01 
    & \cellcolor{gray!30}\textbf{0.99 ± 0.00} 
    & 0.62 ± 0.03 
    & 0.77 ± 0.02 \\

\cmidrule(lr){2-10}
& \textbf{Average (Neural)}    
    & 0.99 ± 0.00
    & 0.99 ± 0.01
    & 0.98 ± 0.00
    & 0.99 ± 0.00
    & 0.83 ± 0.02
    & 0.99 ± 0.00
    & 0.65 ± 0.04
    & 0.79 ± 0.03 \\

\bottomrule

    \end{tabular}
    }
    \label{tbl:ml-detection-augmention}
\end{table*}

\subsection{\UP{Augmented Training with Inlined Functions}}
\label{app:augmented}

\PP{\UP{Experiments and Results}}
\UP{
We perform augmented training with inlined functions 
for a malware detection task (T3) as a potential defense.
We generate new extreme-inlining variants and
add them to the training dataset
that consists of 159 benign samples, 
50 Mirai and 25 Gafgyt samples.
Note that we use a different extreme inlining 
recipe for augmentation 
(\autoref{tab_extreme_inlining_recipe})
to ensure unseen patterns for testing.
The preliminary results in~\autoref{tbl:ml-detection-augmention} 
demonstrate two messages. 
First, the model's performance (\eg F1)
increases in neural networks such as CNN and DNN
after augmented training.
Second, non-neural models, such as
logistic regression and random forest, do
not benefit from augmented training.
}

\begin{figure}[t!]
    \centering
    \resizebox{0.85\linewidth}{!}{%
    \includegraphics[width=0.99\linewidth]{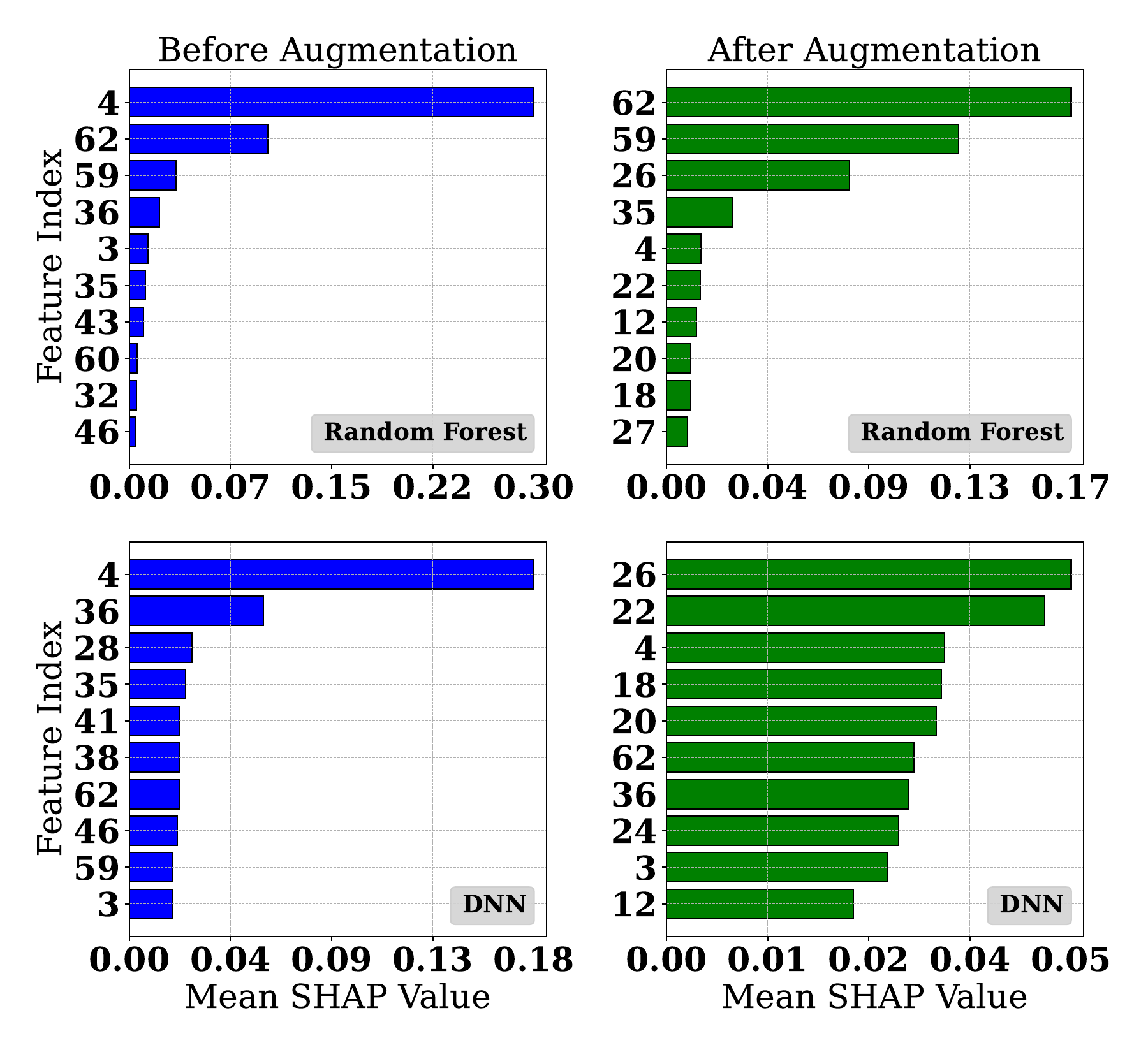}
    }
    \caption{
        \UP{
        The top 10 most influential 
        features (\autoref{tbl:feature_list}) based on 
        mean absolute SHAP values for a 
        traditional ML model (\eg random forest) and 
        a neural network model (\eg DNN) before 
        and after augmentation 
        with extreme inlining. 
        A higher SHAP score denotes a 
        greater contribution of the feature 
        to the model's decision.
        Although baseline feature rankings 
        differ across models, 
        augmentation reshapes feature 
        importance more strongly 
        DNN
        than 
        random forest,
        suggesting that neural 
        architectures adapt more readily 
        to extreme-inlining.
        }
    }
    \label{fig:shap_analysis}
\end{figure}

\PP{\UP{In-depth Analysis with SHAP}}
\UP{
To assess how augmentation alters 
model behavior, 
we apply SHapley Additive exPlanations 
(SHAP)~\cite{lundberg2017unified}
to measure feature contributions across 
ten iterations, comparing
the mean absolute SHAP values before and 
after augmentation.
It is noted that we attempt to 
understand how each feature 
contributes to a model decision, 
as Explainable AI has 
a fidelity issue: \ie 
explanations approximate a decision 
process, but they do not perfectly 
reflect it.
\autoref{fig:shap_analysis}
depicts the Top 10 
contributing features for 
two representative models:
random
forest from a non-neural network
and 
DNN from 
a neural network.
Extreme inlining impacts the entire 
feature space, modifying
not only instruction frequencies but also
control-flow and call-graph
characteristics.
Our findings show that, after augmented 
training, 
neural model 
(DNN) 
exhibit a more pronounced shift 
in feature importance
compared to the 
traditional model 
(random forest)
indicating that augmentation has a more 
substantial effect on neural architectures.
Interestingly, across all models without 
augmentation training, 
the decisions are dominated by Feature 4 
(\ie number of unknown instructions). 
Our further analysis reveals that 
\cc{rep stosq, qword ptr [rdi], rax} 
frequently appears in malware samples, 
which likely accounts for the strong 
contribution of this feature to detection.
}

\begin{figure*}[t!]
    \centering
    \resizebox{\linewidth}{!}{%
    
    \includegraphics[width=\linewidth]{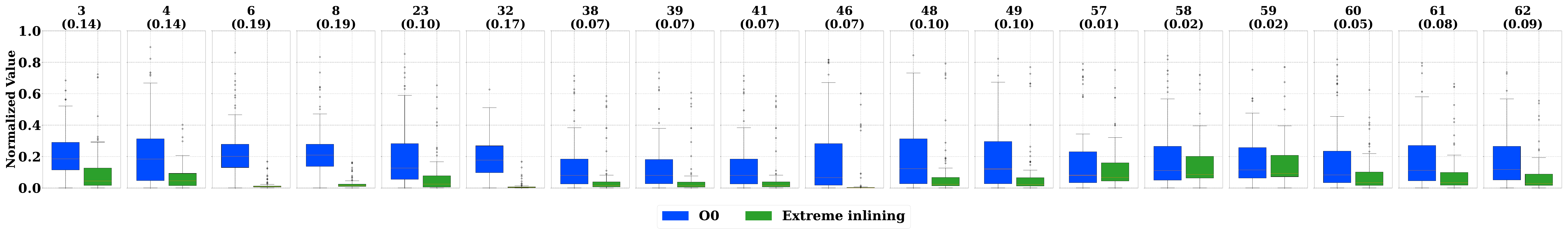}
    }
    \caption{
    Comparison of selective static features
    between little inlining (\cc{-O0}) and
    extreme inlining (\eg compiler options:
    \cc{-O3 -flto=full} 
    \cc{-inline-threshold=200000}).
    The features include
    instructions (Features $3, 4, 6, 8, 23, 32$), 
    control flow graphs (Features $38, 39, 41, 46, 48, 49$), and call graphs (Features $57, 58, 59, 60, 61, 62$).
    The numbers on top denote the features
    and the gap between the two means (parenthesis)
    (\autoref{tbl:feature_list} in Appendix).
    This example illustrates
    the significant gaps in the mean values,
    which can threaten the robustness
    of ML-based models.
    Note that we normalize all values
    for concise comparison.
    See Section~\ref{ss:static_feat} in detail.
    }
    \label{fig:instruction_cfg_cg}
\end{figure*}

\section{Artifact Appendix}

\subsection{Abstract}

Function inlining is a common compiler optimization
that replaces a function call with the function’s 
body to improve performance, but it can also alter the
structural properties of binaries
used in machine learning–based security analysis.  
To investigate this phenomenon, our study systematically
examines the security implications of 
inlining on ML-based binary analysis.
This artifact accompanies our paper and provides 
ready-to-use datasets and scripts for 
verifying the main results presented therein.  
It includes the source code and build scripts used to 
explore inlining-related compiler flags 
toward \emph{extreme inlining}, 
modified feature extraction tools, 
runnable ML models, and analysis
scripts for regenerating the main
results reported in the paper.   
Together, these resources enable examiners to
validate our findings and further examine the impact of extreme inlining 
on ML-based security tasks, including
binary code similarity detection (T1), 
function name prediction (T2), 
malware detection (T3), 
malware family prediction (T4), and 
vulnerability detection (T5).

\subsection{Description \& Requirements}\label{as_req}

The provided artifact includes the source code, 
build scripts, curated datasets, and Python
analysis utilities used to reproduce the
experimental results presented in our paper.
The artifact is organized into four main components:

\begin{itemize}[leftmargin=*]
  \item \textbf{Dataset Construction.} 
  Includes the source code and build scripts
  for exploring inlining-related compiler flags under
  diverse configurations toward \emph{extreme inlining}. 
  Although the full dataset construction process
  is provided for completeness, ready-to-use datasets are
  included to allow quick verification without
  the need for recompilation.

  \item \textbf{Feature Extraction.} Contains a modified version of
  \emph{TikNib}, configured to extract static 
  and semantic features used across 
  Tasks~T3 and~T4 in the study.

  \item \textbf{ML Models.} Provides runnable ML
  models and associated datasets used
  for evaluating robustness under extreme inlining,
  across five binary analysis security tasks (T1--T5) and 
  18 representative models.

  \item \textbf{Main Results and Plots.} Includes CSV
  result files and Python scripts to
  regenerate the primary analysis results reported in the paper. 
  A verification script is also provided to
  reproduce these results directly from
  the included datasets.
  
\end{itemize}

\PP{Scope} This artifact focuses on verifying the 
reported findings rather than regenerating 
datasets from scratch. 
For T3 and T4, due to the stochastic nature
of ML models and Monte Carlo cross-validation, 
exact results may vary slightly from those
reported in the paper; however, the overall 
trends remain consistent.

\PP{Security, privacy, and ethical concerns}
The provided scripts operate entirely on 
open-source data and do not
perform any destructive or privacy-sensitive actions. 
Binaries generated from the proprietary 
SPEC CPU2006 benchmark
are not shared due to
licensing restrictions.
Therefore, no security, privacy, 
or ethical concerns apply.

\subsubsection{How to access}
The artifacts are available open-source 
on Zenodo~\footnote{\url{https://doi.org/10.5281/zenodo.17759528}}.

\subsubsection{Hardware dependencies}
Experiments were conducted on a workstation running 
Ubuntu 20.04 (64-bit) with an Intel Xeon Gold 5218R @ 3.00 GHz 
CPU, 512 GB RAM, and two NVIDIA RTX A6000 GPUs.

\subsubsection{Software dependencies}
The artifact requires the following software environment: 
LLVM/Clang~14.0.0, Python~3.8 or newer 
(with \texttt{pip}~$\ge$~23.0 and \texttt{Conda}~$\ge$~23.7.4), 
Docker~20.10.22 or newer, and IDA~Pro~8.2 or~8.3.

\subsubsection{Benchmarks}

The experiments rely on multiple datasets (See \autoref{tab:binary_dataset}) 
and benchmark 
implementations used across the evaluated
ML-assisted binary analysis tasks. 
Specifically, we use the BCSD benchmark~\cite{marcelli2022machine} 
for T1 and T5, 
including models such as Asm2Vec,
Gemini,
Trex,
SAFE,
and 
BinShot.
For T2, we retrain 
AsmDepictor~\cite{asmdepictor} using our LLVM-compiled dataset. 
For the malware-related tasks (T3 and T4), 
we curated datasets from different sources. 
For T3, we used samples collected from VirusShare~\cite{virusshare}. 
For T4, we curated the dataset from Alrawi~\etal~\cite{alrwai_circle},
whose artifacts are publicly 
available at~\footnote{\url{https://badthings.info/\#bins}}.
The corresponding models include four traditional 
ML classifiers built with Scikit-learn~\cite{ScikitLearn} and
deep learning models re-implemented from
Abusnaina \etal~\cite{Abusnaina}. 
For T5, we use the BCSD
benchmark along with publicly available 
vulnerable firmware corpora~\cite{netgear_tplink_firmware}.

\subsection{Artifact Installation \& Configuration}

\subsubsection{Installation}
To install the artifact, download the repository 
from the provided link and navigate to the project root directory. 
All required dependencies are listed in Section~\ref{as_req} and
can be installed using standard package management 
tools such as \texttt{pip} or \texttt{conda}. 
Each directory is self-contained and 
includes a \texttt{START\_EVALUATION.md} file 
and configuration scripts to guide 
installation and execution.
Note that a pre-configured Docker 
image is also provided, alternatively.

For tasks T1, T2, and T5, create the 
provided Conda environment:

\begin{verbatim}
    cd ml-model/t1_t2_t5
    conda env create -f environment.yml
    conda activate inline_ae
\end{verbatim}

For tasks T3 and T4, using the Conda environment is optional; 
dependencies are available in the root directory:
\begin{verbatim}
    pip install -r requirements.txt
\end{verbatim}

\subsubsection{Basic Test}
To verify that the artifact is correctly installed
and functional, execute any of the provided task scripts. 
For example for T1:

\begin{verbatim}
    cd ml-model/t1_t2_t5/T1_bcsd
    bash run.sh
\end{verbatim}

\subsection{Experiment Workflow}

The artifact is organized into modular components, 
each corresponding to a stage in the experimental workflow. 
The overall process can be summarized as follows: 
(i)~dataset construction under diverse inlining
configurations using the provided
compilation sweep scripts;
(ii)~feature extraction with the 
modified \texttt{TikNib} framework~\cite{tiknib};
(iii)~training and evaluation of machine
learning models across 
five security tasks (T1--T5); and
(iv)~analysis and visualization of the impact of 
inlining on binaries and their associated statistical features, which
can be regenerated using the provided
analysis scripts.

\subsection{Major Claims}

\begin{itemize}[] %
     \item (C1): The artifact provides the 
    complete source code, build scripts,
    and curated datasets used to evaluate the impact
    of function inlining under diverse compiler configurations. 
    These resources enable transparent verification of
    the dataset generation process and reproducibility
    of the experimental setup described in 
    Section~\ref{s:eval}.

    \item (C2): The provided ML models and feature
    extraction tools reproduce all key evaluations across the
    five ML-based security tasks (T1--T5), supporting the
    paper’s findings on the sensitivity of model robustness
    to extreme inlining, as discussed in Section~\ref{ss:eval2}.

    \item (C3): The included analysis scripts regenerate all 
    primary quantitative results and plots (Figures~7--14), 
    illustrating the inlining trends across configurations and 
    the impact of extreme inlining on statistical features 
    compared to the non-inlined baseline, as detailed in Section~\ref{eval1}.
\end{itemize}

\subsection{Evaluation}

\begin{itemize}[] %
 \item (E1): \textbf{Dataset Construction and Compilation Sweep} [15 person-minutes + several compute-hours]  
Run the provided dataset construction scripts to reproduce 
the compilation sweep under diverse inlining configurations. 
The build framework systematically explores optimization
and hidden inlining flags defined in \texttt{config.yaml}. 
Depending on the selected projects and increments, each dataset 
build may take several hours on a multi-core system. 
For validation purposes, reviewers may execute a reduced
configuration (\eg a limited subset of projects or thresholds) to
confirm functionality without requiring
full-scale recompilation.

\item (E2): \textbf{Feature Extraction and Visualization} [10 person-minutes + moderate compute time]  
Execute the modified \texttt{TikNib} pipeline to 
extract binary-level features and
generate summary statistics or visualizations. 
This step produces the feature representations used 
in training ~T3--T4 
and feature distortions 
(Figures~14). 

\item (E3): \textbf{Model Evaluation Across Tasks} 
[10 person-minutes + compute time depending on task]  
Evaluate the robustness of the ML models 
against extreme inlining 
using
the provided scripts in
the \texttt{ml-model/} directory. 
Approximate running times per
task are as follows:  
T1 and T5 — approximately 2~hours,  
T2 — about 1~hour,  
T3 — around 10~minutes, and  
T4 — about 20~minutes.  

\end{itemize}

Successful completion of Experiments~(E1)--(E3) reproduces 
the main findings of the paper, validating the
reproducibility and completeness 
of the provided artifact.

\subsection{Notes on Reusability}

The provided artifact is designed for reuse 
and extension in future research on extreme inlining, 
binary analysis, and ml–based security. 
Each component—dataset construction, 
feature extraction, model evaluation, 
and analysis—is modular and independently executable. 
By adjusting the configuration files 
(\eg, \texttt{config.yaml}) and following the detailed instructions
in the accompanying \texttt{README.md} files, users can
reproduce our experiments or adapt them to new datasets, 
compilers, or 
model architectures.  
For convenience, we also provide a prebuilt 
Docker image for reproducing 
the main experiments.
As a final note, the modular design 
also allows researchers to
explore more extreme inlining degrees 
could be explored automatically through 
our tuning strategy, providing 
an avenue for future work on 
adversarial code transformation.

\end{document}